\documentclass[twocolumn,nofootinbib,floatfix,
superscriptaddress]{revtex4}        %start writting 27 June 12%
\usepackage{graphicx}
\usepackage{epsfig}
\usepackage{bm}
\usepackage[T1]{fontenc}
\usepackage[latin9]{inputenc}
\usepackage{amssymb}
\usepackage{float}
\usepackage{amsmath}
\usepackage{dcolumn}
\usepackage{cancel}
\usepackage[colorlinks]{hyperref}
\usepackage[usenames,dvipsnames]{color}
\hypersetup{
     breaklinks=true,
    pdfstartview={FitH},    % fits the width of the page to the window
    colorlinks=true,       % false: boxed links; true: colored links
    linkcolor=blue,          % color of internal links
    citecolor=red,        % color of links to bibliography
    filecolor=magenta,      % color of file links
    urlcolor=blue,           % color of external links
    anchorcolor=green,      % Color for anchor text
    linktocpage=true
}

\newcommand{\be}{\begin{equation}}
\newcommand{\ee}{\end{equation}}
\newcommand{\De}{\Delta}

\begin{document}

\title{Cosmic evolution and thermal stability of Barrow holographic dark energy in nonflat Friedmann-Robertson-Walker Universe}

\author{G.~G.~Luciano}
\email{gluciano@sa.infn.it}
\affiliation{Applied Physics Section of Environmental Science Department, Universitat de Lleida, Av. Jaume
II, 69, 25001 Lleida, Catalonia, Spain}
\affiliation{INFN, Sezione di Napoli, Gruppo collegato di Salerno, Via Giovanni Paolo II, 132 I-84084 Fisciano (SA), Italy}

\date{\today}

\begin{abstract}
We study the cosmological evolution of a nonflat
Friedmann-Robertson-Walker Universe filled 
by pressureless dark matter and Barrow Holographic
Dark Energy (BHDE). The latter is a dark
energy model based on the holographic principle with 
Barrow entropy instead of the 
standard Bekenstein-Hawking one. By assuming the
apparent horizon of the Universe as IR cutoff, 
we explore both the cases where a mutual interaction 
between the dark components of the cosmos is absent/present. 
We analyze the behavior of various model parameters, such
as the BHDE density parameter, the equation of state
parameter, the deceleration parameter, the jerk parameter and
the square of sound speed. We also comment
on the observational consistency of our predictions, 
showing that the interacting model turns out to be favored
by recent experimental constraints from Planck+WP+BAO, SNIa+CMB+LSS  and Union2 SNIa joint analysis over the non-interacting one.
The thermal stability of our framework 
is finally discussed by demanding the positivity of the
heat capacities and compressibilities of the cosmic fluid. 
\end{abstract}
 \maketitle

\section{Introduction}
\label{Intro}
Despite the recent progress at both theoretical
and experimental level, there are still several
open problems in modern Cosmology~\cite{Open}. 
Among these, the issue of the observed
accelerating expansion of the Universe
has been getting greater attention in the last decades~\cite{Supern,Supernbis,Supernter,Supernquar,Supernquin}, 
providing one of the most exciting challenges 
to the Standard Model of particle physics and Cosmology.
A widely accepted paradigm to explain this
phenomenon is the existence of a mysterious
form of energy with negative pressure - the Dark Energy (DE) - which would
dominate the energy-matter budget of the present Universe. 

An interesting scenario for the quantitative description of DE
is Holographic Dark Energy (HDE)~\cite{Cohen:1998zx,Horava:2000tb,Thomas:2002pq,Li:2004rb,Hsu:2004ri,Huang:2004ai,Wang:2005ph,Setare:2006sv,Granda,Sheykhi:2011cn,Bamba:2012cp,Ghaffari:2014pxa,Wang:2016och,Moradpour:2020dfm}, which proves to be in good agreement with observations~\cite{Zhang,Li,Zhang2,Lu}.
A crucial ingredient of this framework is the application of the holographic principle at cosmological scales. According
to this recipe, the total entropy content of the Universe
should be proportional to the area of its two-dimensional boundary, 
similarly to the Bekenstein-Hawking entropy-area law for 
black holes. Recently, however, it has been argued
that quantum-gravitational effects may induce nontrivial
fractal features on the black hole structure, leading
to a deformed horizon entropy in the form~\cite{BarrowBH}
\be
\label{BE}
S_\De\,\propto\,A^{1+\Delta/2}\,, 
\ee
where $A$ is the area enclosed by the horizon surface. 
Corrections to the standard entropy-area law
are parameterized by the exponent $0\leq\Delta\leq1$, 
where $\Delta=0$ recovers the Bekenstein-Hawking limit, 
while $\Delta=1$ corresponds to the maximal entropy deformation. 
It is worth noting that upper bounds on $\Delta$ have been derived in
different contexts in~\cite{Anagnostopoulos:2020ctz,Leon:2021wyx,Barrow:2020kug,Jusufi:2021fek,Dabrowski:2020atl,Saridakis:2020cqq,Luciano:2022pzg,Vagnozzi:2022moj}.

Based on the gravity-thermodynamic conjecture~\cite{GTC}, 
Eq.~\eqref{BE} is commonly adopted also in Cosmology~\cite{BarSar,Sheykhi:2021fwh,Adhikary:2021xym,Nojiri:2021jxf,LucLiu,GhafLuc} 
to derive modified Friedmann equations 
from the first law of thermodynamics on the apparent horizon of a Friedmann-Robertson-Walker (FRW) Universe. 
The ensuing framework is referred to as Barrow Holographic Dark Energy (BHDE) and exhibits a
richer phenomenology compared to HDE, allowing us 
to predict effects beyond the domain
of the Standard Model of Cosmology for certain values of $\Delta$ (see, for instance,~\cite{Luciano:2022pzg}). 

In line with the above studies, the properties of some 
DE models have been largely investigated
in~\cite{Radicella,Mimoso,MoradInt,Santos,Barboza,Bhandari}
at thermodynamic level. 
Specifically, in~\cite{Barboza} the limits imposed
by thermodynamics on a DE fluid have been
analyzed by considering the heat capacities
and compressibilities. On the other hand, in~\cite{Bhandari}
the thermal stability criteria have been explored for interacting DE. A similar analysis has been carried out in~\cite{Abdollah}
for the case of Tsallis Holographic Dark Energy~\cite{Tavayef:2018xwx,Saridakis:2018unr,Nojiri:2019skr,DAgostino,LucGine}, which
is a generalization of HDE based on the use
of nonextensive Tsallis entropy for the horizon 
degrees of freedom of Universe~\cite{Cirto}.
All these studies suggest the existence of an intimate relation
between the cosmic expansion of the Universe, 
the thermodynamic stability conditions and DE candidates, 
which deserves careful attention.

Starting from the above premises, in this work we examine
the cosmic evolution of a nonflat
FRW Universe filled
by pressureless dark matter and BHDE.  By considering
the apparent horizon as IR cutoff and assuming 
a mutual interaction between the dark sectors of the cosmos, 
we analyze the behavior of various model parameters, 
such as the BHDE density parameter, the equation of state
parameter, the deceleration parameter, the jerk parameter and
the square of sound speed. 
We then discuss 
the thermal stability of our model 
by demanding the positivity of the heat capacities and compressibilities of the cosmic fluid. 

Some comments are in order here to motivate
our study: first, we emphasize that in the absence
of a fully consistent theory of quantum gravity, 
BHDE represents a first, naive approach to accommodate 
quantum-gravity corrections within the HDE framework
based on simple physical principles~\cite{BarSar}. 
Furthermore, the usage of the apparent horizon as IR cutoff
is substantiated by the fact that, unlike the event horizon, 
this boundary is consistent with the generalized second law of thermodynamics~\cite{She1,She2}. Last but not least, 
although it is commonly believed that inflation washed out
curvature inhomogeneities in the early Universe, observational evidences
suggest that there is still room for a small but non-negligible
spatial curvature~\cite{Chernin,Ellis,PLB,Divale}.
This is why we consider a nonflat FRW Universe
as background for our analysis. 

The remainder of the work is organized as follows: in the next section
we briefly review the theoretical framework of BHDE. 
The evolution of the characteristic model parameters 
is studied in Sec.~\ref{Evolution} for both
the noninteracting and interacting cases. 
In Sec.~\ref{Thermal} we discuss the thermal stability of BHDE, 
while conclusions and outlook are finally summarized in
Sec.~\ref{Conc}. Throughout the work, we use natural units
$\hbar=c=k_B=G=1$.

\section{Barrow Holographic Dark Energy: a review}
\label{InterBHDE}
Let us start by briefly reviewing the main features
of BHDE. It is well-known that the standard HDE is given by the inequality~\cite{Wang:2016och}
\be
\rho_{D} L^4 \leq S\,,
\ee
with the further assumption $S\propto A\propto L^2$, 
where $\rho_D$ denotes the dark energy density, while $L$ and $S$ 
are the horizon length and entropy, respectively. 

On the other hand, the usage of Barrow entropy~\eqref{BE}
leads to~\cite{BarSar}
\be
\label{rdeB}
\rho_D=C L^{\Delta-2}\,,
\ee
where $C$ is an unknown parameter with dimensions $[L]^{-2-\Delta}$.
Notice that, for $\Delta=0$, the above relation
reduces to the usual HDE, provided that $C=3c^2m_p^2$, where $m_p$ is the reduced Planck mass and $c$
the model parameter. 

Now, in a nonflat FRW Universe containing pressureless dark matter and BHDE, the first Friedmann equation takes the form
\be
\label{FFE}
H^2+\frac{k}{a^2}=\frac{8\pi}{3}\left(\rho_M+\rho_D\right),
\ee
where $a$ is the scale factor, $H\equiv \dot a/a$ is the Hubble
parameter and the dot denotes derivative with respect to the cosmic
time $t$. 

We also assume that the apparent horizon of radius
\be
\label{radius}
\tilde r_A=\frac{1}{\sqrt{H^2+k/a^2}}
\ee 
acts as IR cutoff\footnote{There is no universal consensus on the choice of the IR cutoff. Following~\cite{Tavayef:2018xwx,IRCUT,IRCUTbis}, here we resort to Eq.~\eqref{radius}. However, 
other possible choices are the particle horizon, the future event horizon, the GO cutoff~\cite{GO} or combination thereof.  Given the degree of arbitrariness in the selection of the most reliable description of dark energy, we leave the analysis of BHDE with different IR cutoffs for future work.}~\cite{IRCUT,IRCUTbis}, so that Eq.~\eqref{rdeB}
can be cast as
\be
\label{IRcut}
\rho_D=C \tilde r_A^{\Delta-2}\,.
\ee
The spatial curvature $k$ takes value $+1,0$ or $-1$, depending on whether the shape of the Universe is a closed $3$-sphere, flat space or an open $3$-hyperboloid, respectively. Consistently with recent observations~\cite{Chernin,Ellis,PLB,Divale}, henceforth we consider a closed Universe with a small positive curvature.

By introducing the fractional energy densities
\begin{subequations}
\label{fract}
\be
\Omega_D=\frac{8\pi\rho_D}{3H^2}\,,
\ee
\be
\Omega_M=\frac{8\pi\rho_M}{3H^2}\,,
\ee
\be
\Omega_k=\frac{k}{H^2a^2}\,,
\ee
\end{subequations}
for BHDE, DM and curvature terms, respectively, 
Eq.~\eqref{FFE} can be rearranged as
\be
\label{rearrange}
1=\Omega_D+\Omega_M-\Omega_k\,,
\ee
which can be further manipulated with
the definition of the parameter $r=\rho_M/\rho_D=\Omega_M/\Omega_D$ to give
\be
r=-1+\frac{1}{\Omega_D}\left(1+\Omega_k\right). 
\ee

If now assume the existence of a mutual interaction $Q$
between the dark sectors of the cosmos, the conservation
equations for BHDE and DM can be coupled as
\begin{eqnarray}
\label{rhod}
&\dot\rho_D+3H\rho_D\left(1+\omega_D\right)=-Q\,,\\[2mm]
&\dot\rho_M+3H\rho_M=Q\,,
\label{rhom}
\end{eqnarray}
where $\omega_D\equiv p_D/\rho_D$ is the equation of state (EoS) parameter
of BHDE and $p_D$ its pressure. 

Following~\cite{He}, 
we consider the rate $Q$ of energy exchange between
BHDE and DM in the form
\be
\label{Q}
Q=3b^2H\left(\rho_D+\rho_M\right),
\ee 
where $b^2$ is a dimensionless constant
that quantifies the coupling between the dark sectors. Hence, we have $Q>0$, which implies that a net flux of energy
flows from BHDE to DM. This is consistent with
the second law of thermodynamics and Le Chatelier-Braun
principle~\cite{Q2}.

\section{Cosmological evolution of Barrow Holographic Dark Energy in nonflat Universe}
\label{Evolution}
Let us now examine the time evolution of the above model.
To this aim, we insert the time derivative of Friedmann equation~\eqref{FFE} into Eq.~\eqref{rhod} and combine the resulting
expression with Eqs.~\eqref{rearrange} and~\eqref{rhom}.
After some algebra, we get
\be
\label{dotH}
\frac{\dot H}{H^2}=\Omega_k-\frac{3\Omega_D}{2}\left(1+r+\omega_D\right).
\ee
From this equation, we can straightforwardly derive the deceleration parameter as
\begin{eqnarray}
\nonumber
q&=&-\frac{\ddot a}{aH^2}=-1-\frac{\dot H}{H^2}\\[2mm]
&=&-1-\Omega_k+\frac{3\Omega_D}{2}\left(1+r+\omega_D\right).
\label{qpar}
\end{eqnarray}
It is easy to see that
positive values of $q$ indicate a decelerated expansion 
of the Universe ($\ddot a<0$), while 
negative values correspond to an accelerated phase ($\ddot a>0$). 

In a similar fashion, by combining the time derivative of Eq.~\eqref{IRcut}
with~\eqref{radius} and~\eqref{dotH}, we obtain
\be
\label{dotrho}
\dot\rho_D=\left(2-\Delta\right)\rho_D\,H\left[\Omega_k-\frac{3\Omega_D}{2}\left(1+r+\omega_D\right)\right], 
\ee
while the use of Eqs.~\eqref{fract} and~\eqref{dotH} 
leads to
\begin{eqnarray}
\nonumber
\label{OmegaD}
\Omega_D'&=&-\Omega_D\left[\frac{1}{2}\left(2-\Delta\right)\left(3+\Omega_k+3\Omega_D\,\omega_D\right)\right.\\[2mm]
&&+\,2\Omega_k-3\Omega_D\left(1+r+\omega_D\right)
\bigg].
\end{eqnarray}
Here we have used the standard notation 
$\Omega'_D\equiv\frac{d\Omega_D}{d(\log a)}$.

A relevant quantity to establish whether
the model is classically stable against small perturbations
is the square of sound speed, which is defined by
\be
\label{vsq}
v_s^2=\frac{dp_D}{d\rho_D}=\omega_D+\dot\omega_D \frac{\rho_D}{\dot\rho_D}\,.
\ee
We remark that classical stability requirement is satisfied, 
provided that $v_s^2>0$. Indeed, for a density perturbation,
positive values of the squared sound speed correspond to
a regular propagation mode. Vice versa, negative
values imply that the perturbation equation becomes
an irregular wave equation~\cite{Pert1,Pert2}. 
As a consequence, with a density perturbation,  
the negative squared speed indicates an escalating mode.
In other terms, when the density perturbation increases, 
the pressure decreases, favoring the development 
of an instability~\cite{Pert1,Pert2}.

In what follows we study the evolution trajectories
of the model parameters introduced above. We analyze
the noninteracting ($b^2=0$) and interacting ($b^2\neq0$) cases
separately, discussing their consistency with
observations.

\subsection{Noninteracting model}
In the case where there is no interaction between the dark
sectors of the cosmos, insertion of Eq.~\eqref{dotrho} into~\eqref{rhod} 
allows us to infer the following expression for the BHDE EoS parameter 
\be
\label{EoSpa}
\omega_D=-\frac{3+\left(\frac{\Delta}{2}-1\right)\left(\Omega_k+3\right)}
{3\left[1+\left(\frac{\Delta}{2}-1\right)\Omega_D\right]}\,.
\ee
In turn, from Eq.~\eqref{OmegaD} we get
\be
\label{numerical}
\Omega_D'=\frac{\Omega_D\,\Delta}{2}\,\frac{3+\Omega_k-3\Omega_D}{1+\left(\frac{\Delta}{2}-1\right)\Omega_D}\,.
\ee
 
\begin{figure}[t]
\begin{center}
\includegraphics[width=8.5cm]{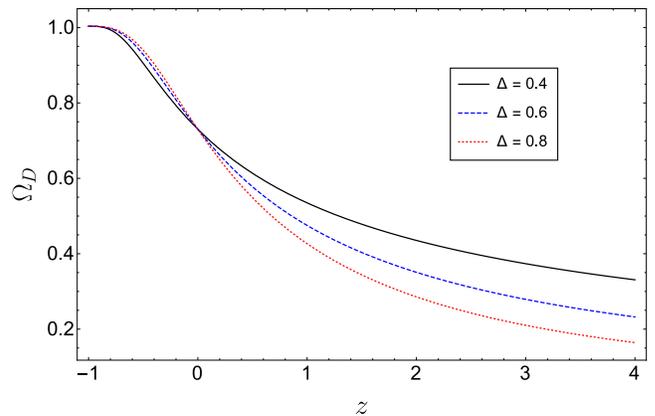}
\caption{The evolution of $\Omega_D$ versus $z$ for different values
of $\Delta$. We have set $\Omega_k=0.01$ and $\Omega_D^0=0.73$ as initial condition.}
\label{Fig1}
\end{center}
\end{figure}

\begin{figure}[t]
\begin{center}
\includegraphics[width=8.5cm]{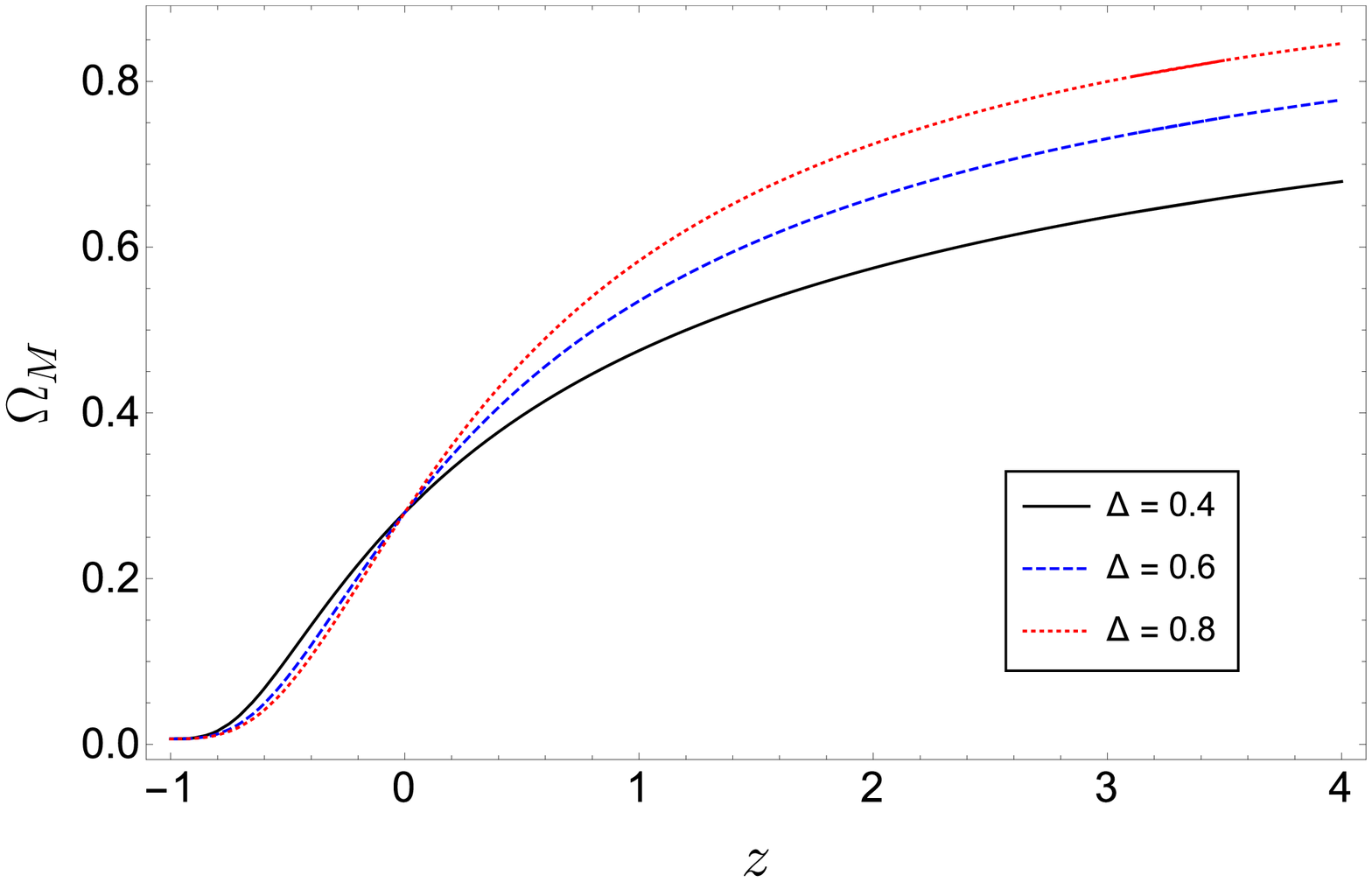}
\caption{The evolution of $\Omega_M$ versus $z$ for different values
of $\Delta$. We have set $\Omega_k=0.01$ and $\Omega_D^0=0.73$ as initial condition.}
\label{Fig1bis}
\end{center}
\end{figure}

\begin{figure}[t]
\begin{center}
\includegraphics[width=8.5cm]{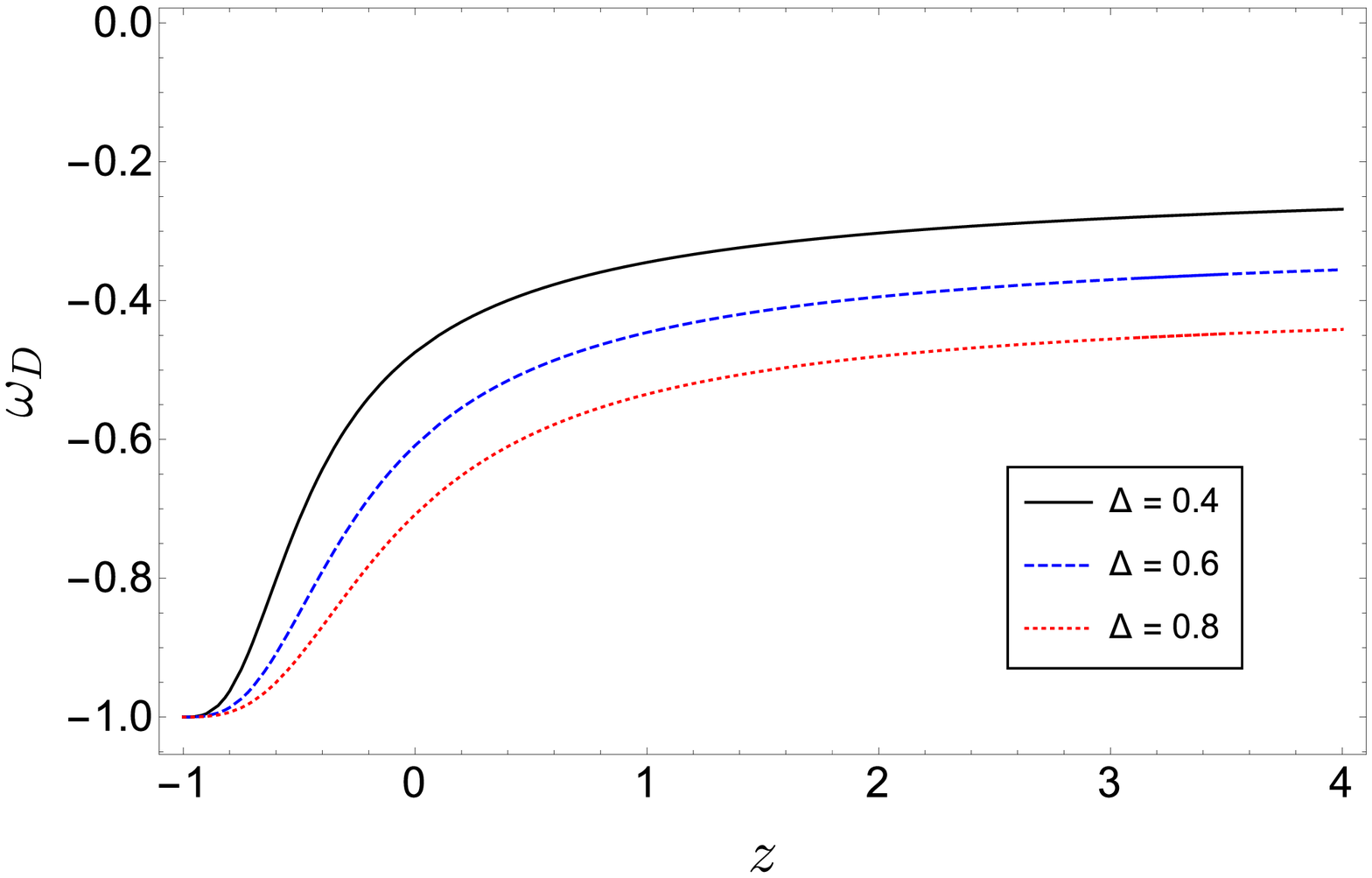}
\caption{The evolution of $\omega_D$ versus $z$ for different values
of $\Delta$. We have set $\Omega_k=0.01$ and $\Omega_D^0=0.73$ as initial condition.}
\label{Fig2}
\end{center}
\end{figure}

\begin{figure}[t]
\begin{center}
\includegraphics[width=8.5cm]{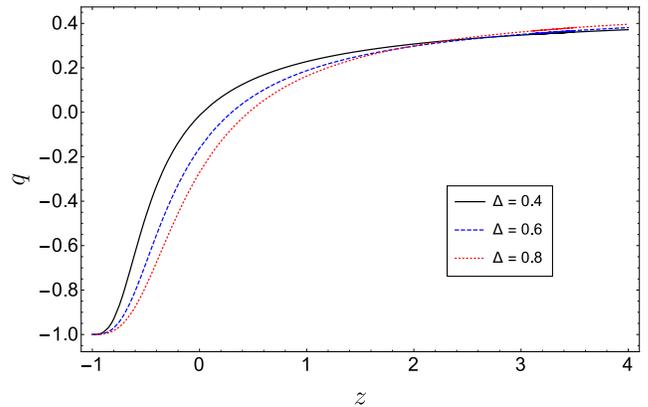}
\caption{The evolution of $q$ versus $z$ for different values
of $\Delta$. We have set $\Omega_k=0.01$ and $\Omega_D^0=0.73$ as initial condition.}
\label{Fig3}
\end{center}
\end{figure}

To determine the behavior of BHDE, 
we solve this equation numerically. 
The evolution of $\Omega_D$ and $\Omega_M$
versus the redshift $z=\left(1-a\right)/a$
is plotted in Fig.~\ref{Fig1} and Fig.~\ref{Fig1bis} for different values of $\Delta$, respectively. From Fig.~\ref{Fig1}
it is evident that $\Omega_D$ increases monotonically
going from early times to the far future. This indicates the evolution of the present model from an initial dark matter dominated phase 
toward a completely dark energy
dominated epoch. The complementary behavior
is exhibited by $\Omega_M$ (see Fig.~\ref{Fig1bis}).

The dynamics of EoS parameter is shown in Fig.~\ref{Fig2}. 
As we can see, $\omega_D$ lies in the quintessence regime
($-1<\omega_D<-1/3$) at present (i.e. $z=0$) and tends
to the cosmological constant behavior $\omega_D=-1$ 
in the far future ($z\rightarrow-1$), regardless of $\Delta$.\footnote{Notice that any comparison with the results of~\cite{Adhikary:2021xym} is prevented
by the fact that a different IR cutoff is used in that case.
} Specifically, for the current value
of the EoS parameter the present model predicts
$\omega_{D}^0\in[-0.72,-0.47]$ for the considered values of $\Delta$.
Notice that this range is consistent with neither the recent 
observational constraints obtained from Planck+WP+Union 2.1
($-1.26<\omega_0<-0.92$) and  Planck+WP+BAO ($-1.38<\omega_0<-0.89$) data, nor with WMAP+eCMB+BAO+H0 ($-1.162<\omega_0<-0.983$) measurements~\cite{Planck},
thus requiring some proper amendment
to this model.

For completeness, let us analyze the evolution of the deceleration 
parameter. By plugging Eqs.~\eqref{EoSpa} and~\eqref{numerical}
into~\eqref{qpar}, this takes the form
\be
\label{qdece}
q=\frac{1+\Omega_k-\left(1+\Delta\right)\Omega_D}
{2+\left(\Delta-2\right)\Omega_D}\,.
\ee
The behavior of $q$ is plotted in 
Fig.~\ref{Fig3} for different values of $\Delta$. 
As explained above, 
our model predicts the successive sequence of an early
matter dominated era with a decelerated expansion ($q>0$), followed by a late time DE dominated
epoch with an accelerated phase ($q<0$), 
In particular, the present framework is capable of explaining the observed accelerated phase of the Universe today ($z=0$), in contrast to the standard Holographic Dark Energy model. However, quantitatively speaking
the redshift $z_t$ at which
the transition from the decelerated to accelerated Universe occurs, i.e. $q(z_t)=0$,  
lies within the interval 
$z_t\in[0.02,0.46]$ 
for the considered values of the model parameters. This
range does not overlap with the observational constraint obtained
via SNIa+BAO/CMB ($z_t=0.72\pm0.05$)~\cite{MamonEPJC2},   SNIa+CMB ($z_t=0.57\pm0.07$)~\cite{Alam} and SNIa+CMB+LSS joint analysis ($z_t=0.61$)~\cite{MamonMPA}.
Also, the value of the deceleration parameter
for the current epoch is estimated as $q_0\in[-0.26,-0.01]$,
to be compared with the experimental outcome $q_0=-0.64\pm0.22$ from Union2 SNIa data~\cite{WuHu}.
These results provide further evidence
of the observational inconsistency of the non-interacting model.

\begin{figure}[t]
\begin{center}
\includegraphics[width=8.5cm]{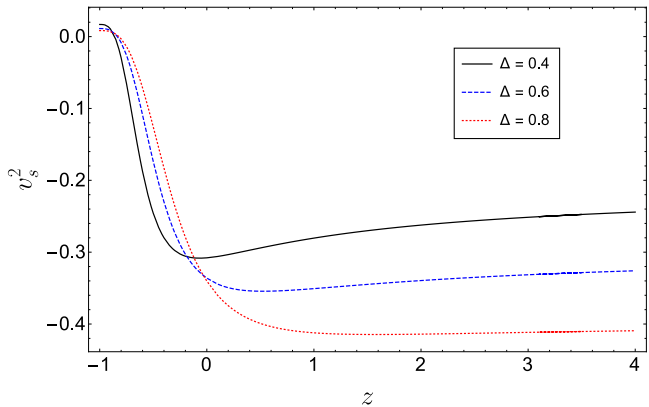}
\caption{The evolution of $v_s^2$ versus $z$ for different values
of $\Delta$. We have set $\Omega_k=0.01$ and $\Omega_D^0=0.73$ as initial condition.}
\label{Fig5}
\end{center}
\end{figure}

From Eq.~\eqref{vsq}, we can finally derive
the square of sound speed as
\be
\label{vsquare}
v_s^2=\frac{\left[\frac{3\Delta}{2}+\left(\frac{\Delta}{2}-1\right)\Omega_k\right]\left(\Omega_D-1\right)}
{3\left[1-\left(1-\frac{\Delta}{2}\right)\Omega_D\right]^2}\,.
\ee
This is plotted as a function of the redshift $z$
in Fig.~\ref{Fig5}. Interestingly enough, we see that
$v_s^2$ takes negative values at early and present times, 
while it tends to positive values in the far future.
In other terms, the present model evolves towards
a classically stable configuration. A similar result
has been exhibited in~\cite{hybrid} for the case
of BHDE with hybrid expansion law. 

\begin{figure}[t]
\begin{center}
\includegraphics[width=8.5cm]{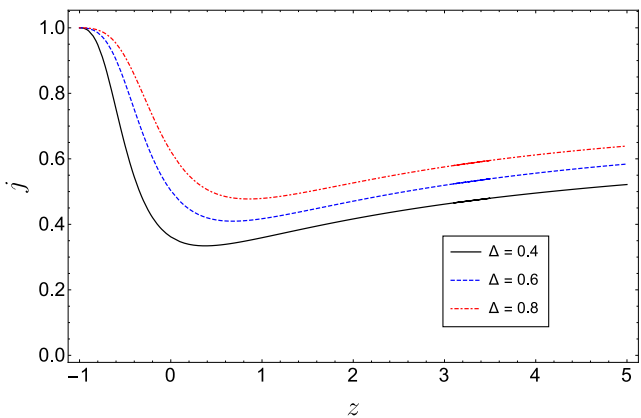}
\caption{The evolution of $j$ versus $z$ for different values
of $\Delta$. We have set $\Omega_k=0.01$ and $\Omega_D^0=0.73$ as initial condition.}
\label{jerk1}
\end{center}
\end{figure}

Figure~\ref{jerk1} finally shows the behavior
of the jerk parameter, which is given by the 
third derivative of the scale factor
respect to the cosmic time, i.e.~\cite{Rapetti}
\be
\label{jerk}
j\,=\,\frac{1}{aH^3}\frac{d^3a}{dt^3}=q\left(2q+1\right)+\left(1+z\right)\frac{dq}{dz}\,.
\ee
This quantity allows us to quantify departures from $\Lambda$CDM model, which
is characterized by $j=1$. 
From Fig.~\ref{jerk1}, we find that $j>0$ for any
redshift and approaches unity in the far future (i.e. $z\rightarrow-1$), where our model fits with 
$\Lambda$CDM. Also, we have
$j_0\in[0.36,0.62]$ in the present epoch for the considered
values of model parameters. 
Notice that deviations of $j_0$ from the $\Lambda$CDM value need attention as the real cause behind the cosmic acceleration is still unknown~\cite{Bambabis}.

\subsection{Interacting model}
We now extend the above analysis to the case
where the interaction~\eqref{Q} between the dark sectors of the cosmos is taken into account. 
By repeating the same computations as above, we
derive the following generalized expression for the EoS parameter
\be
\omega_D=-\frac{3+\left(\frac{\Delta}{2}-1\right)\left(\Omega_k+3\right)+3b^2\left(1+r\right)}
{3\left[1+\left(\frac{\Delta}{2}-1\right)\Omega_D\right]}\,.
\ee

Similarly,  we get
\begin{eqnarray}
\Omega_D'&=&\frac{\Omega_D\,\Delta}{2}\,\frac{3+\Omega_k-3\Omega_D-3b^2\left(1+\Omega_k\right)}{1+\left(\frac{\Delta}{2}-1\right)\Omega_D}\,,\\[3mm]
\label{qnonin}
q&=&\frac{1+\Omega_k-\left(1+\Delta\right)\Omega_D-3b^2\left(1+\Omega_k\right)}
{2+\left(\Delta-2\right)\Omega_D}\,,\\[3mm]
\nonumber
v_s^2&=&\frac{\left[\frac{3\Delta}{2}+\left(\frac{\Delta}{2}-1\right)\Omega_k\right]\left(\Omega_D-1\right)}
{3\left[1-\left(1-\frac{\Delta}{2}\right)\Omega_D\right]^2}\\[2mm]
&&+\frac{b^2\left(1+\Omega_k\right)\left[1+\frac{\Delta}{2}+\frac{2}{\left(\Delta-2\right)\Omega_D}\right]}{\left[1-\left(1-\frac{\Delta}{2}\right)\Omega_D\right]^2}\,,
\end{eqnarray}
for the BHDE density, the deceleration parameter and
the squared sound speed, respectively. It is straightforward to check that 
these relations correctly reduce to Eqs.~\eqref{EoSpa},~\eqref{numerical},~\eqref{qdece} and~\eqref{vsquare} 
for $b^2=0$.

\begin{figure}[t]
\begin{center}
\includegraphics[width=8.5cm]{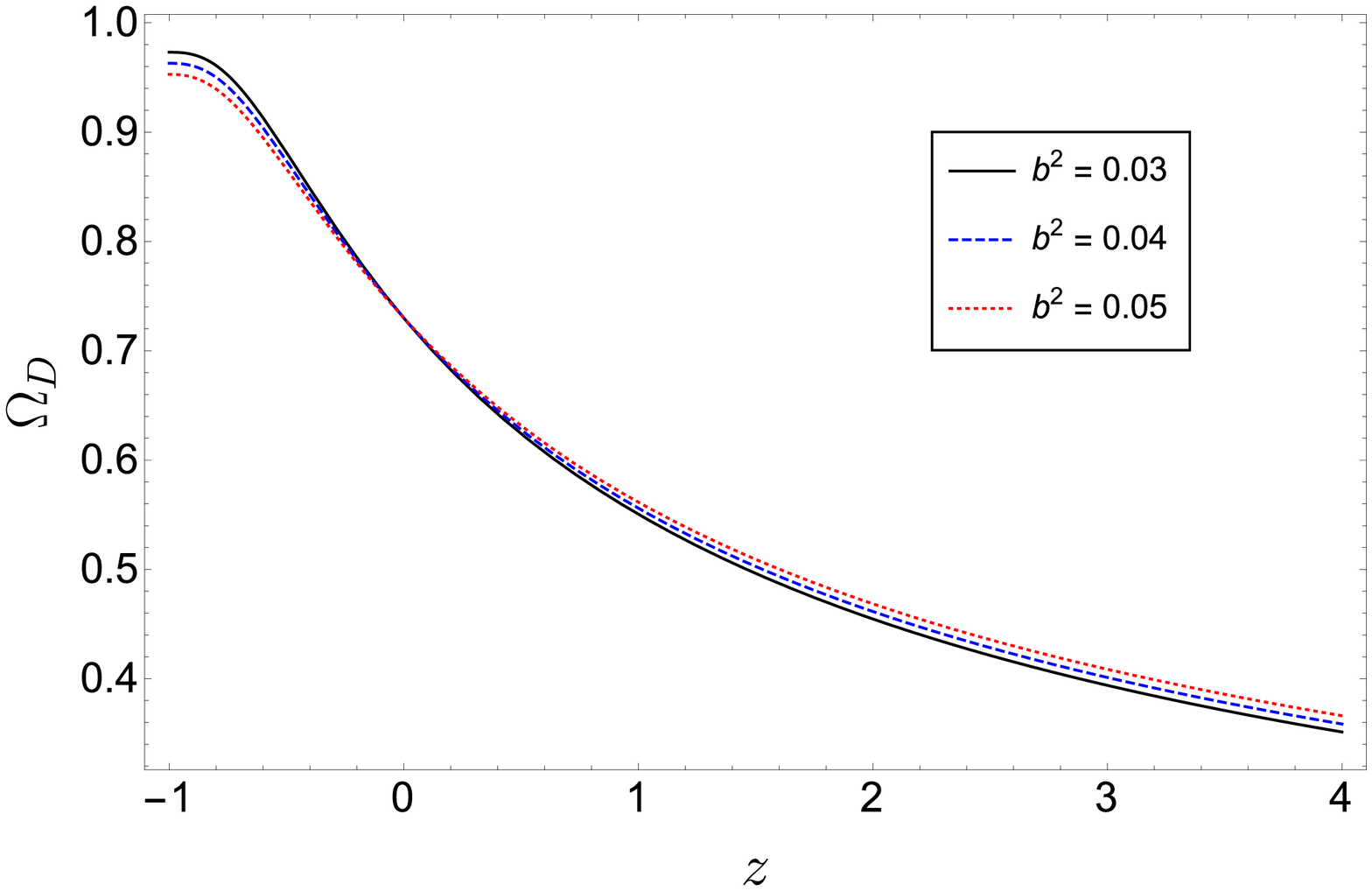}
%\caption{The evolution of $\Omega_D$ versus $z$ for different values
%of $b^2$. We have set $\Omega_k=0.01$, $\Delta=0.4$ and $\Omega_D^0=0.73$ as initial condition.}
%\label{Fig6}
\end{center}
\begin{center}
\includegraphics[width=8.5cm]{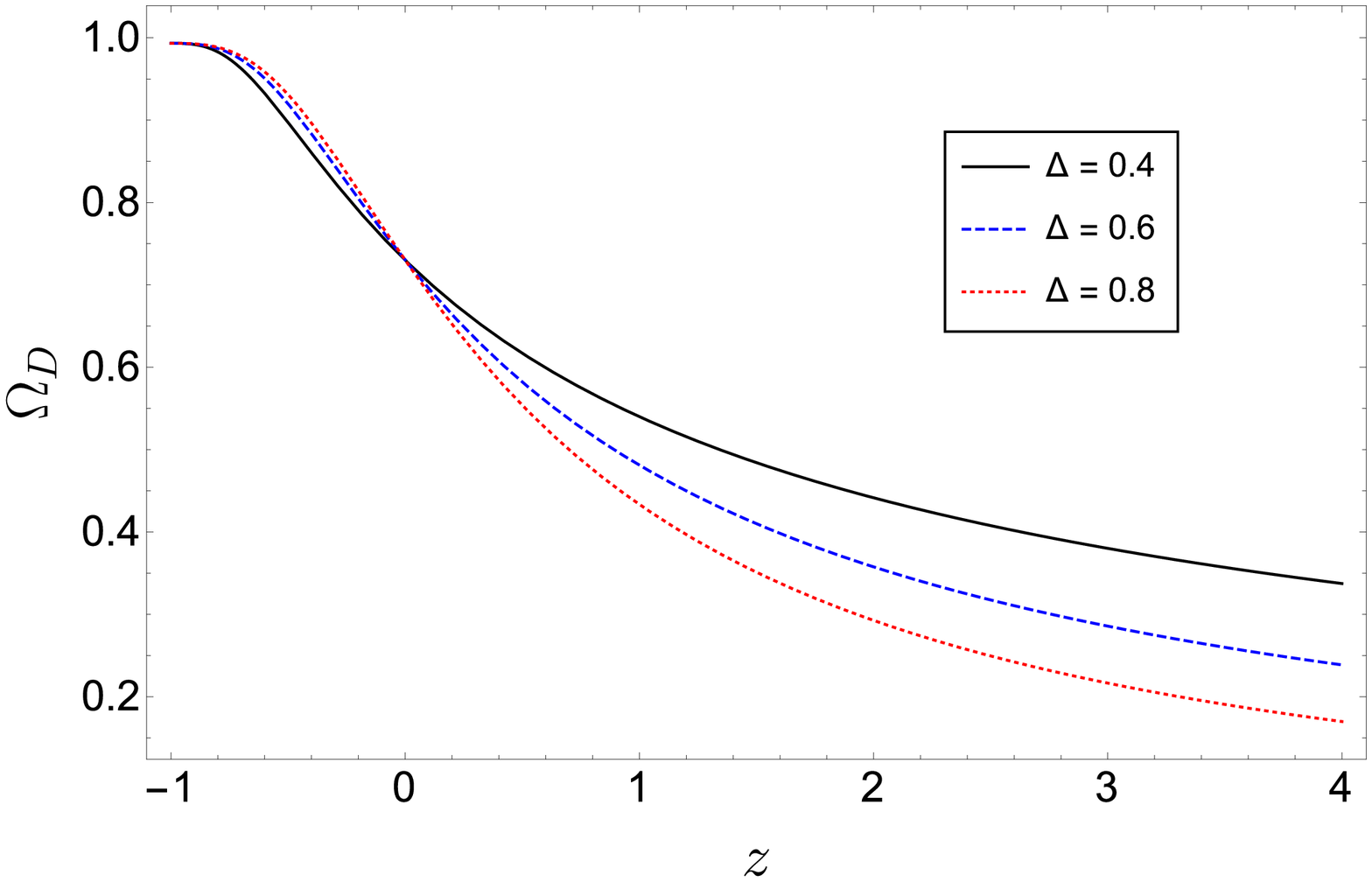}
\caption{The evolution of $\Omega_D$ versus $z$ for different values
of $b^2$ (upper panel) and $\Delta$ (lower panel). We have set $\Omega_k=0.01$ and $\Omega_D^0=0.73$ as initial condition. For the upper plot, we have considered $\Delta=0.4$, while for the lower one $b^2=0.01$.}
\label{Fig6}
\end{center}
\end{figure}

\begin{figure}[t]
\begin{center}
\includegraphics[width=8.5cm]{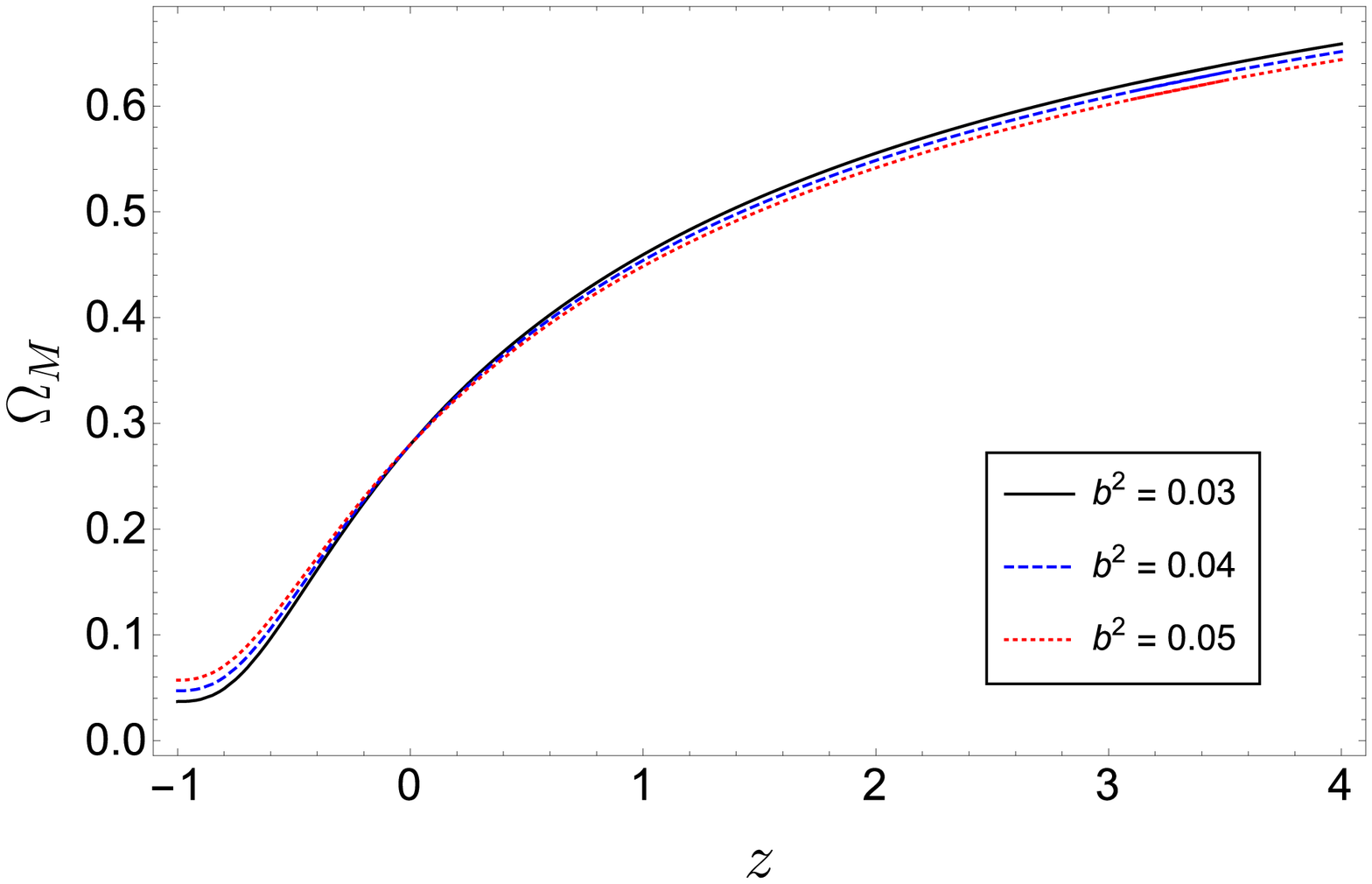}
%\caption{The evolution of $\Omega_D$ versus $z$ for different values
%of $b^2$. We have set $\Omega_k=0.01$, $\Delta=0.4$ and $\Omega_D^0=0.73$ as initial condition.}
%\label{Fig6}
\end{center}
\begin{center}
\includegraphics[width=8.5cm]{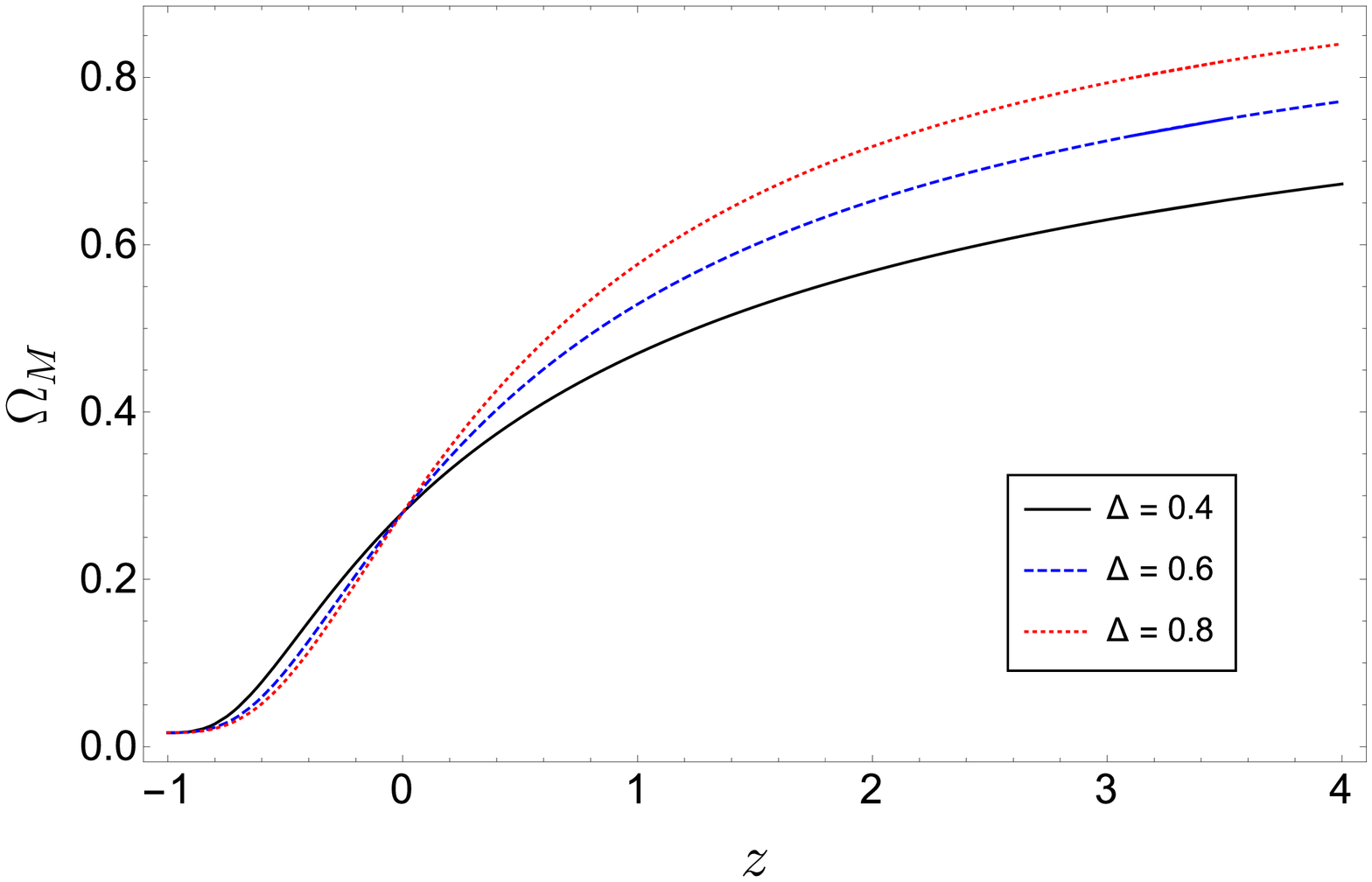}
\caption{The evolution of $\Omega_M$ versus $z$ for different values
of $b^2$ (upper panel) and $\Delta$ (lower panel). We have set $\Omega_k=0.01$ and $\Omega_D^0=0.73$ as initial condition. For the upper plot, we have considered $\Delta=0.4$, while for the lower one $b^2=0.01$.}
\label{Fig7}
\end{center}
\end{figure}

\begin{figure}[t]
\begin{center}
\includegraphics[width=8.5cm]{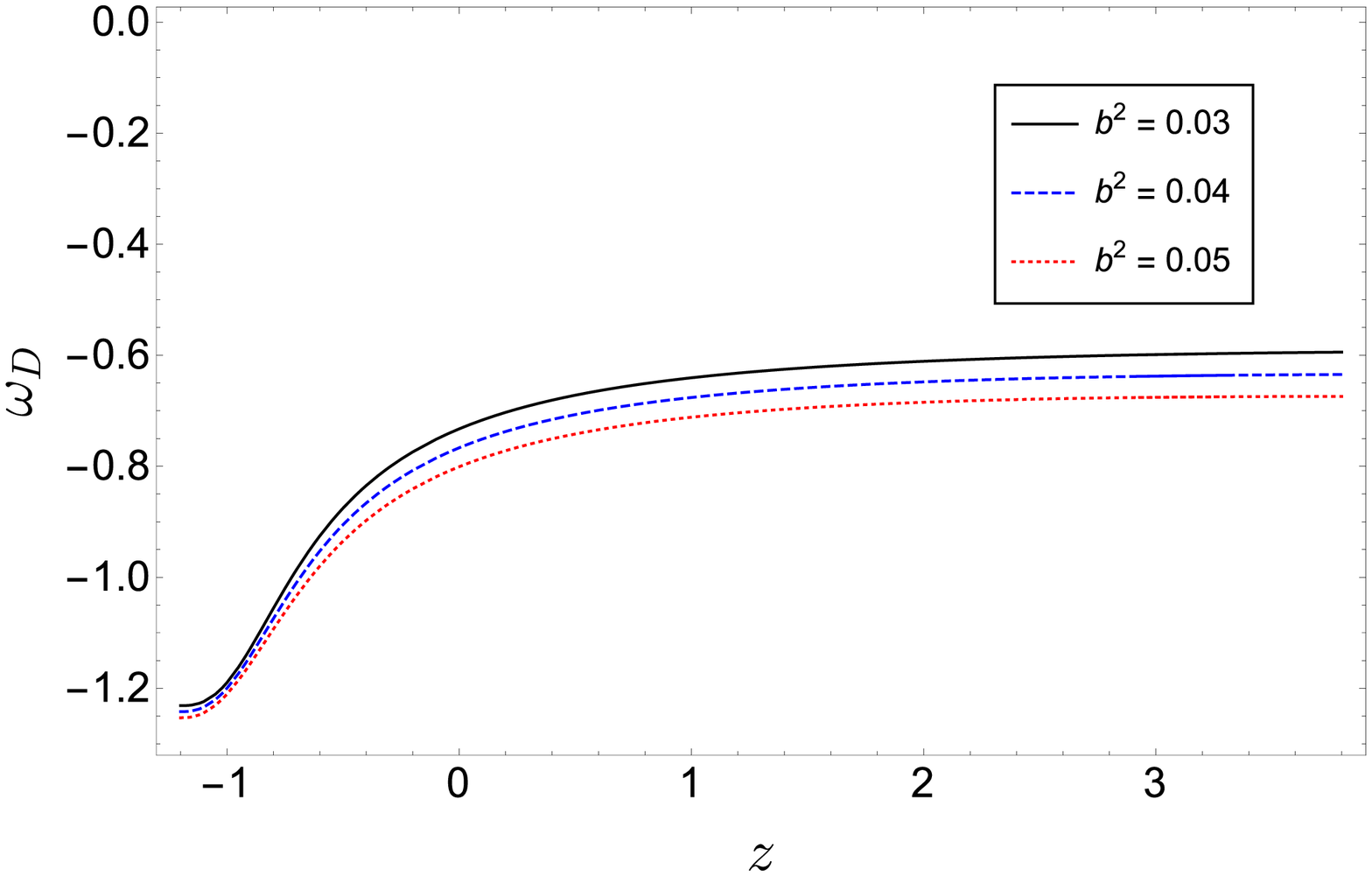}
%\caption{The evolution of $\Omega_D$ versus $z$ for different values
%of $b^2$. We have set $\Omega_k=0.01$, $\Delta=0.4$ and $\Omega_D^0=0.73$ as initial condition.}
%\label{Fig6}
\end{center}
\begin{center}
\includegraphics[width=8.5cm]{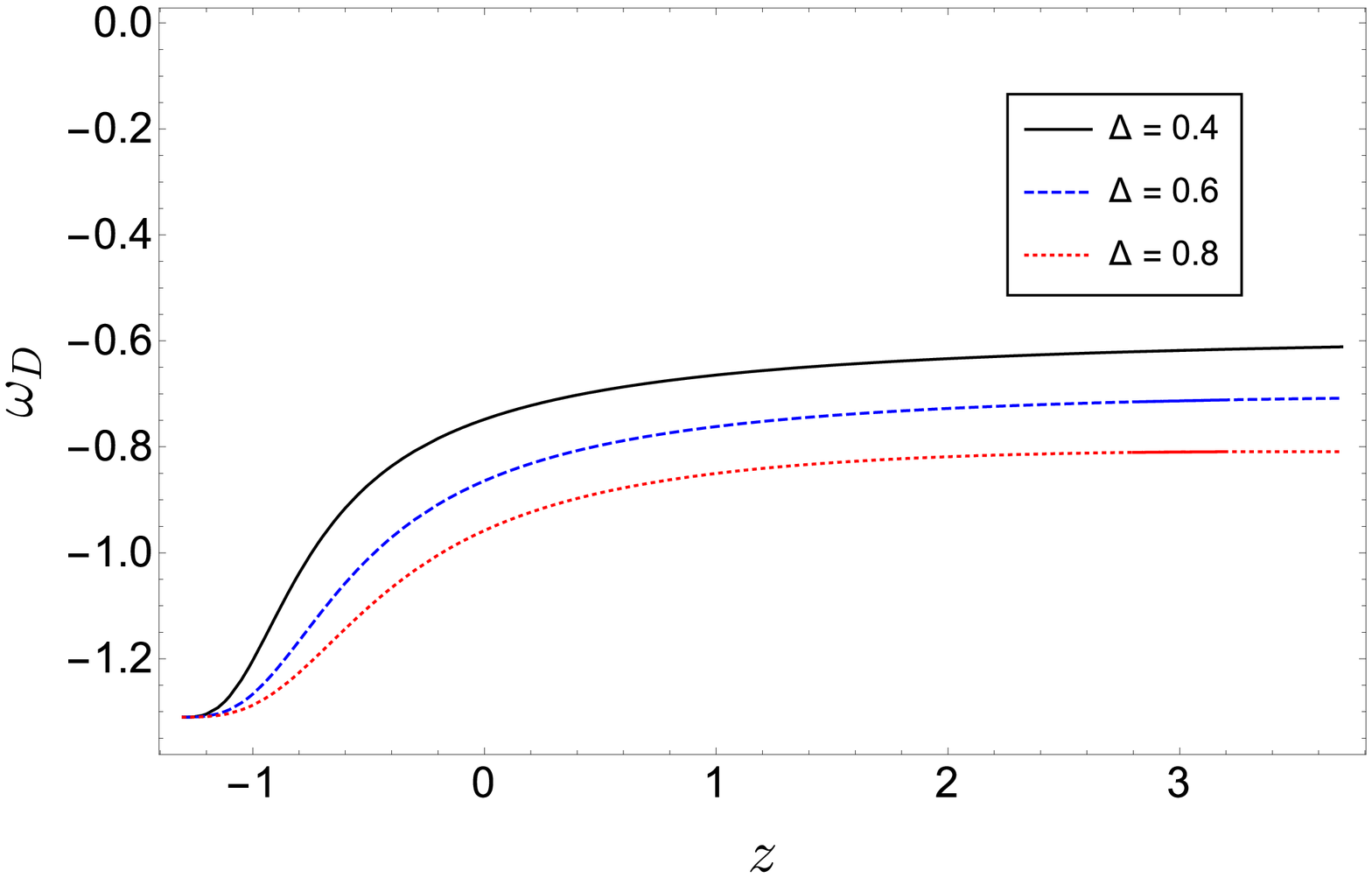}
\caption{The evolution of $\omega_D$ versus $z$ for different values
of $b^2$ (upper panel) and $\Delta$ (lower panel). We have set $\Omega_k=0.01$ and $\Omega_D^0=0.73$ as initial condition. For the upper plot, we have considered $\Delta=0.4$, while for the lower one $b^2=0.01$.}
\label{Fig8}
\end{center}
\end{figure}

Fig.~\ref{Fig6} shows the evolution of $\Omega_D$
versus $z$ for different values of $b^2$ (upper panel) and $\Delta$ (lower panel). On the other hand, in Fig.~\ref{Fig7} we plot 
the behavior of $\Omega_M$. We can see that both these two
densities exhibit
the same profile as for the noninteracting case (see Fig.~\ref{Fig1} and~\ref{Fig1bis} for comparison). Remarkable differences can instead be highlighted from the analysis of the EoS parameter in Fig.~\ref{Fig8}. 
Indeed, in the present study
the phantom line $\omega_D=-1$ is crossed in the
far future. In particular, the higher the coupling, 
the more evident such a behavior
(see upper panel of Fig.~\ref{Fig8}). This
feature is peculiar to this model and has no correspondence in the noninteracting description (see Fig.~\ref{Fig2}). 
To check the observational consistency of the present model, let us compute the current value of the EoS parameter. From the lower panel of Fig.~\ref{Fig8}, we find $\omega_{D}^0\in[-0.96,-0.74]$, which is in agreement with constraints
from Planck+WP+BAO and Planck+WP+Union 2.1 measurements (see the discussion above Eq.~\eqref{qdece}).

\begin{figure}[t]
\begin{center}
\includegraphics[width=8.5cm]{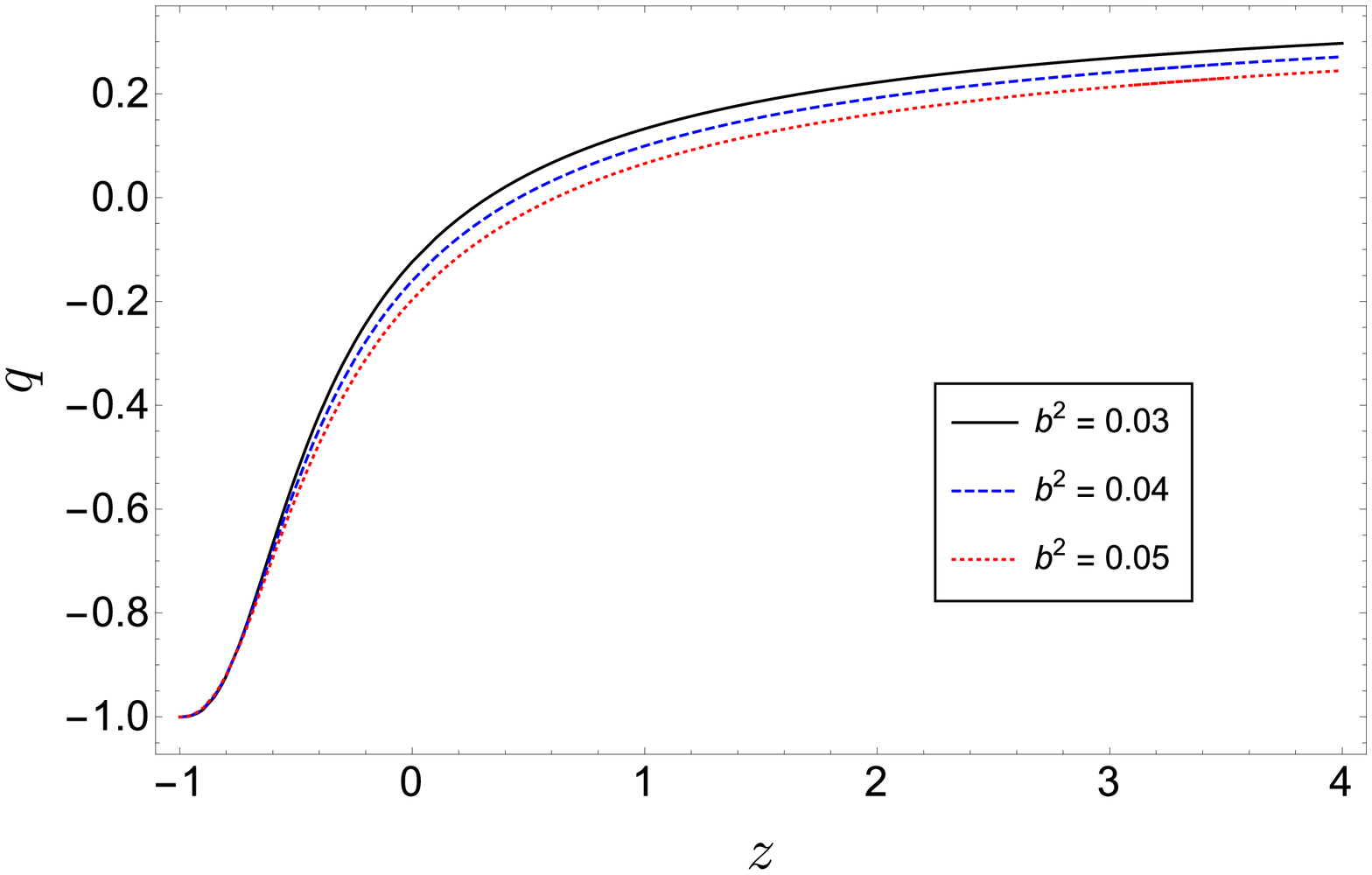}
%\caption{The evolution of $\Omega_D$ versus $z$ for different values
%of $b^2$. We have set $\Omega_k=0.01$, $\Delta=0.4$ and $\Omega_D^0=0.73$ as initial condition.}
%\label{Fig6}
\end{center}
\begin{center}
\includegraphics[width=8.5cm]{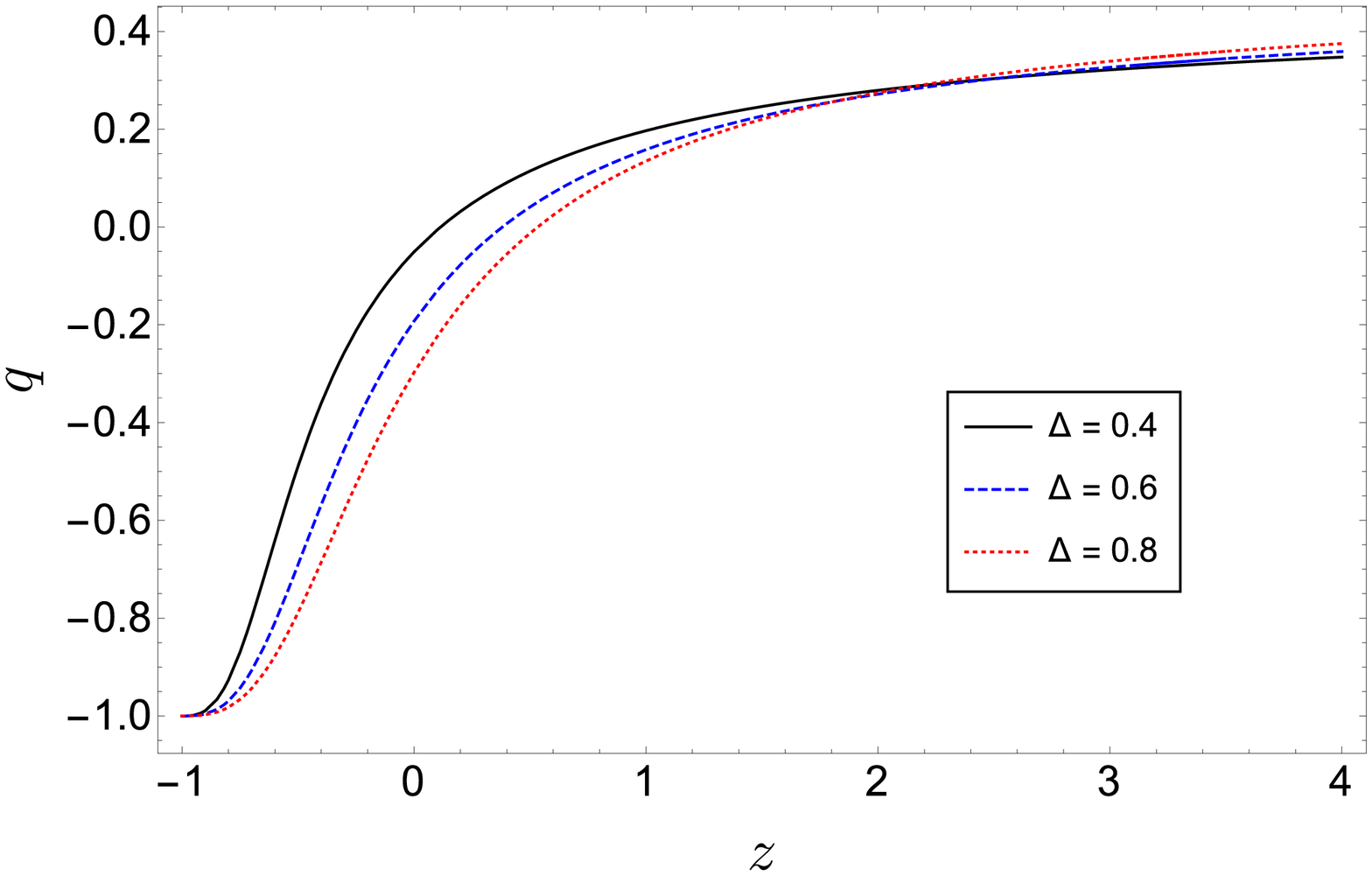}
\caption{The evolution of $q$ versus $z$ for different values
of $b^2$ (upper panel) and $\Delta$ (lower panel). We have set $\Omega_k=0.01$ and $\Omega_D^0=0.73$ as initial condition. For the upper plot, we have considered $\Delta=0.4$, while for the lower one $b^2=0.01$.}
\label{Fig9}
\end{center}
\end{figure}

In Fig.~\ref{Fig9}  we show
the evolution of the deceleration parameter $q$. We see that the interacting BHDE model still predicts
a smooth transition of the Universe from an initial
decelerated phase to a late-time accelerated expansion. 
However, the advantage of this framework over the non-interacting one is that the transition redshift is now estimated as
$z_t\in[0.30,0.64]$, 
which is in good agreement with SNIa+CMB+LSS joint analysis. 
Also, the deceleration parameter for the current epoch
is found to lie in the range $q_0\in[-0.33,-0.03]$ (see the lower panel in Fig.~\ref{Fig9}), the lower bound of which is closer to the experimental value exhibited in~\cite{WuHu}. 
Thus, the present model turns out to be favored by observations.

\begin{figure}[t]
\begin{center}
\includegraphics[width=8.5cm]{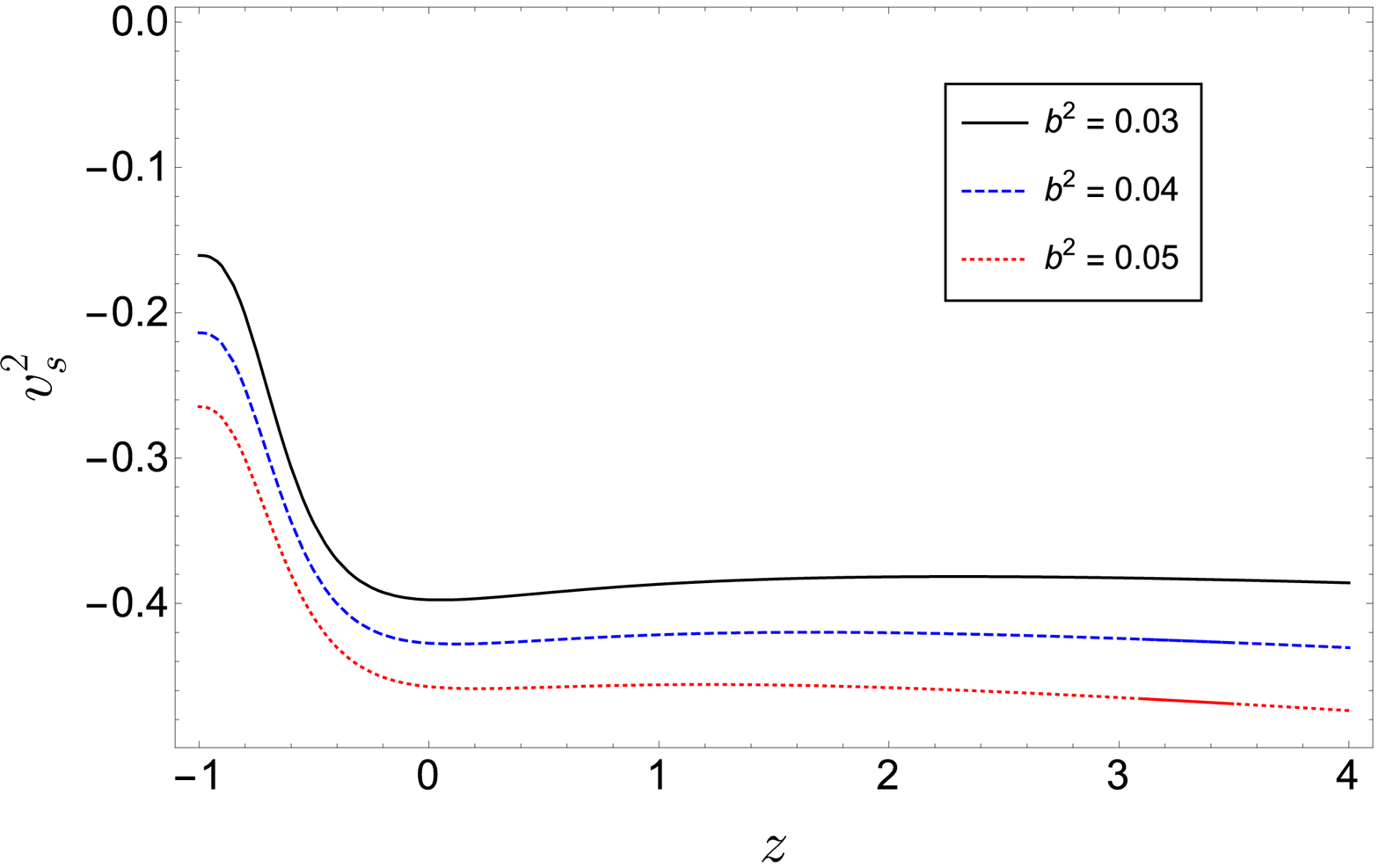}
%\caption{The evolution of $\Omega_D$ versus $z$ for different values
%of $b^2$. We have set $\Omega_k=0.01$, $\Delta=0.4$ and $\Omega_D^0=0.73$ as initial condition.}
%\label{Fig6}
\end{center}
\begin{center}
\includegraphics[width=8.5cm]{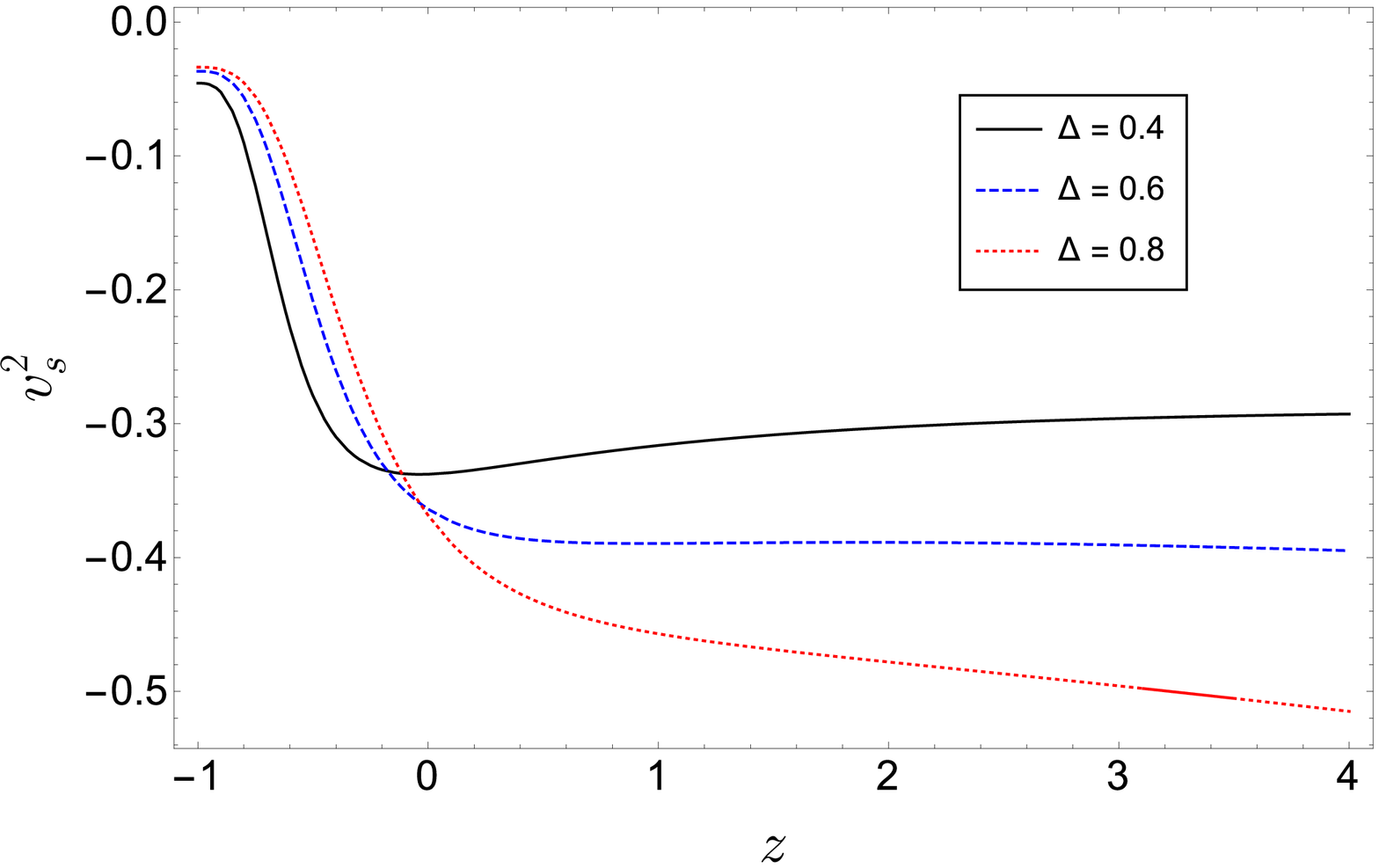}
\caption{The evolution of $v_s^2$ versus $z$ for different values
of $b^2$ (upper panel) and $\Delta$ (lower panel). We have set $\Omega_k=0.01$ and $\Omega_D^0=0.73$ as initial condition. For the upper plot, we have considered $\Delta=0.4$, while for the lower one $b^2=0.01$.}
\label{Fig10}
\end{center}
\end{figure}

The behavior of the squared sound speed is plotted
in Fig.~\ref{Fig10}. By comparison with Fig.~\ref{Fig5}, 
we see that $v_s^2$ is now always negative,  which implies
that interacting BHDE is unstable at classical level against small perturbations.

\begin{figure}[t]
\begin{center}
\includegraphics[width=8.5cm]{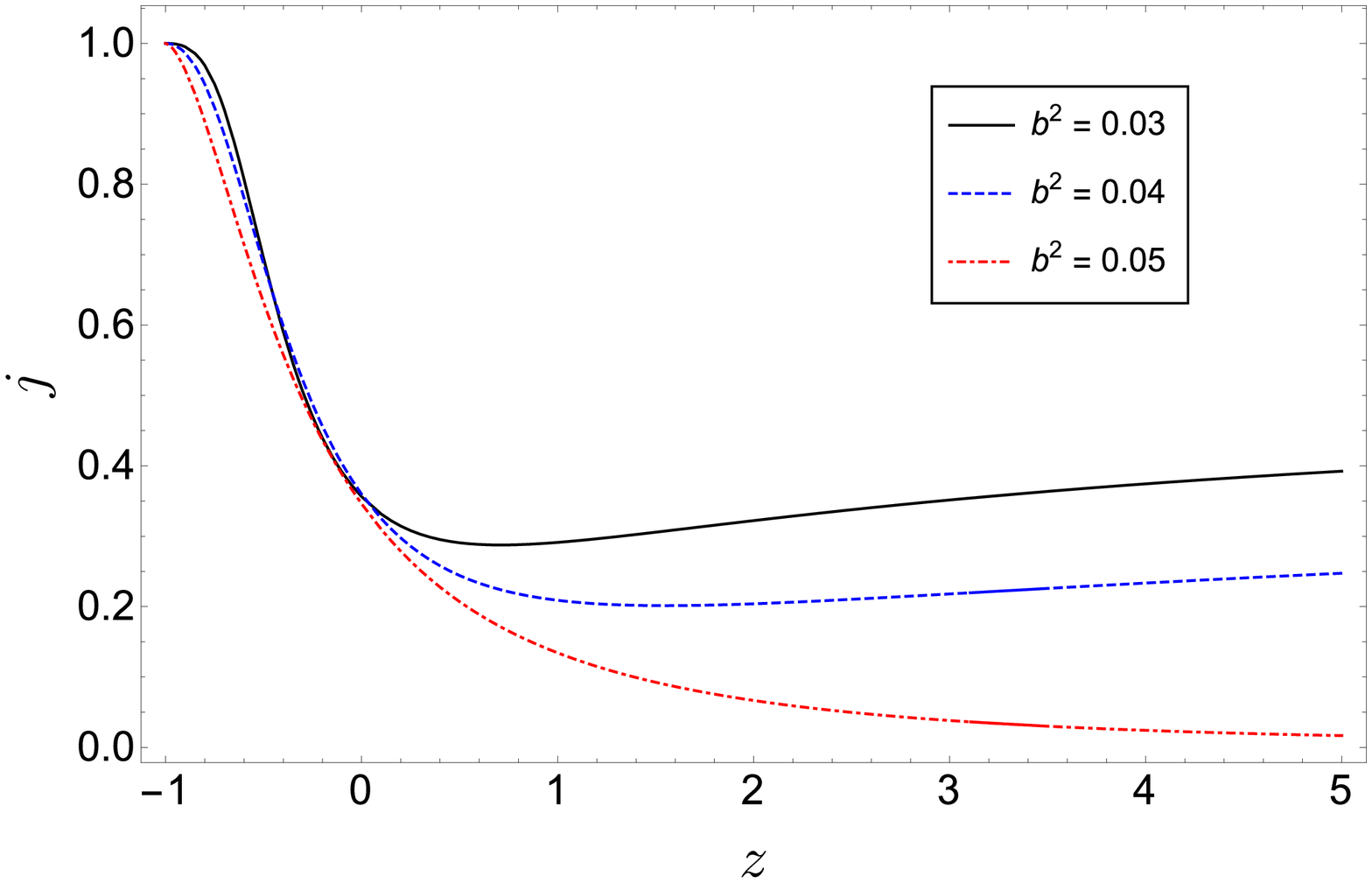}
%\caption{The evolution of $\Omega_D$ versus $z$ for different values
%of $b^2$. We have set $\Omega_k=0.01$, $\Delta=0.4$ and $\Omega_D^0=0.73$ as initial condition.}
%\label{Fig6}
\end{center}
\begin{center}
\includegraphics[width=8.5cm]{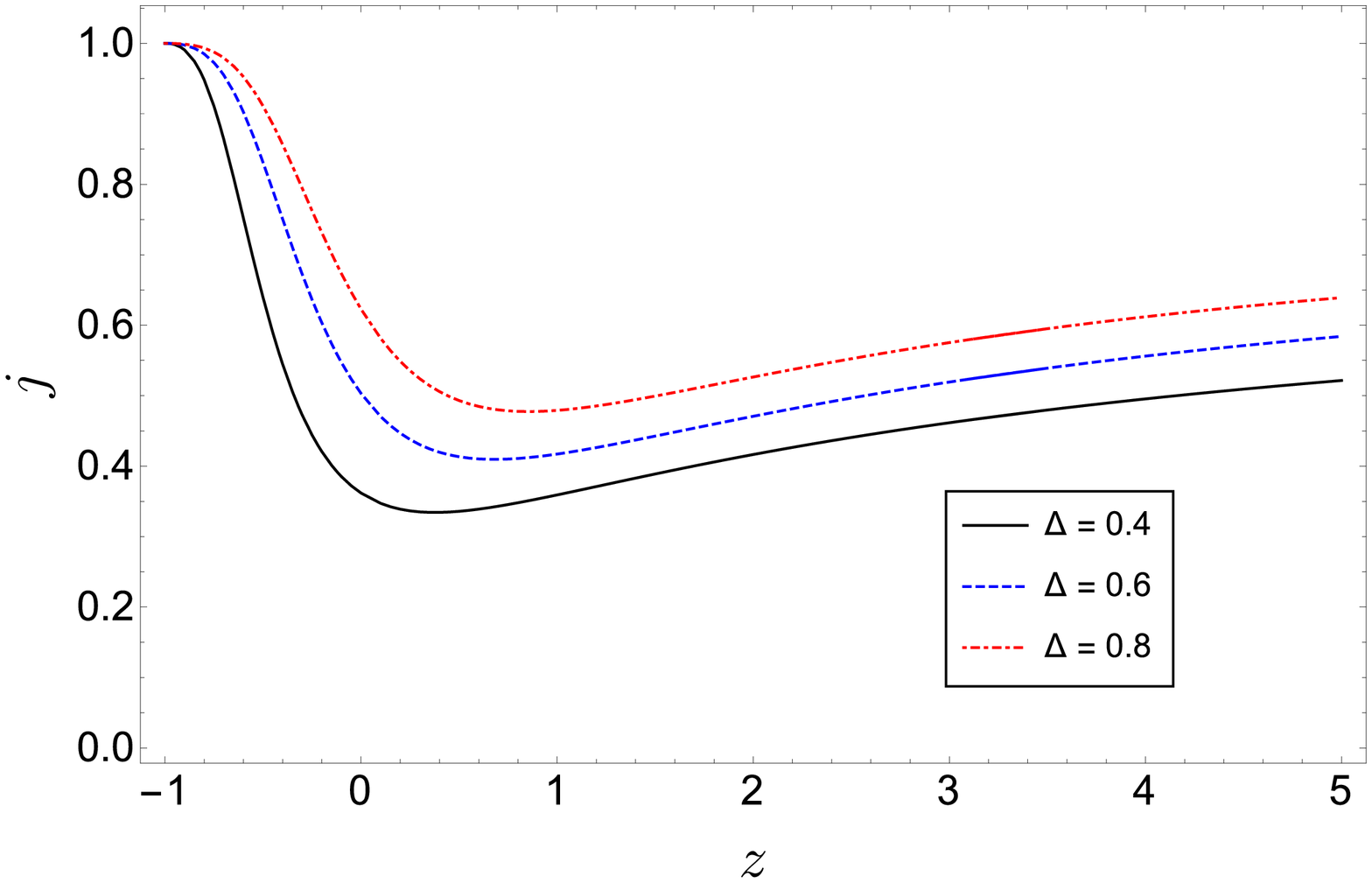}
\caption{The evolution of $j$ versus $z$ for different values
of $b^2$ (upper panel) and $\Delta$ (lower panel). We have set $\Omega_k=0.01$ and $\Omega_D^0=0.73$ as initial condition. For the upper plot, we have considered $\Delta=0.4$, while for the lower one $b^2=0.01$.}
\label{jerk2}
\end{center}
\end{figure}

Figure~\ref{jerk2} shows the jerk parameter~\eqref{jerk} versus the redshift. As in the absence of interactions, we
find that $j>0$ for any redshift, $j_0\in[0.35,0.64]$ (see the lower panel in Fig.~\ref{jerk2}) and $j\rightarrow1$ in the far future (i.e. $z\rightarrow-1$),
thus reproducing $\Lambda$CDM in this limit.

To further investigate the experimental
relevance of the present model and constrain the free parameters $\Delta$, $b$ and $H_0$, let us consider the evolution
of $H(z)$ from Eqs.~\eqref{qpar} and~\eqref{qnonin} for fixed $\Omega_k=0.01$ and $\Omega_D^0=0.73$,  and compare it
with the data points obtained from the 
latest compilation of 57 Hubble's parameter measurements in the range $0.07 \le z \le 2.36$. Such points have been derived via Differential Age (31 points), BAO and other methods (the remaining 26 points) and are listed in Table~\ref{TabI} (see~\cite{Koussour} and references therein for more details on the data set).

Following~\cite{Koussour}, we use the statistical $R^2$-test to find the best fit values of model parameters. This is defined
by 
\be
R^2\,=\,1-\frac{\sum_{i=1}^{57}\left[(H_i)_{ob}-(H_i)_{th}\right]^2}{\sum_{i=1}^{57}\left[(H_i)_{ob}-(H_i)_{mean}\right]^2}\,,
\ee
where $(H_i)_{ob}$ and $(H_i)_{th}$ are the observed and predicted
values of Hubble's parameter, respectively. 
Minimization of deviation of $R^2$ from unity gives the best-fit values $b=0.12$ and $\Delta=0.08$ at $95\%$ confidence level. 
We notice that the obtained estimation
of Barrow parameter fits with
the result $\Delta=0.094^{+0.093}_{-0.101}$ derived in~\cite{Anagnostopoulos:2020ctz} and is very close to
the constraint $\Delta=0.075^{+0.001}_{-0.002}$ of~\cite{Mathew}, 
but it lies outside the intervals $0.005\le\Delta\le0.008$~\cite{Luciano:2022pzg} and  $\Delta\lesssim1.4\times10^{-4}$~\cite{Barrow:2020kug} found via baryogenesis and Big Bang Nucleosynthesis measurements, respectively. This discrepancy might be understood  by assuming a HDE description of the Universe with a running Barrow entropy, as discussed in~\cite{Dige}.

\begin{figure}[t]
\begin{center}
\includegraphics[width=8.5cm]{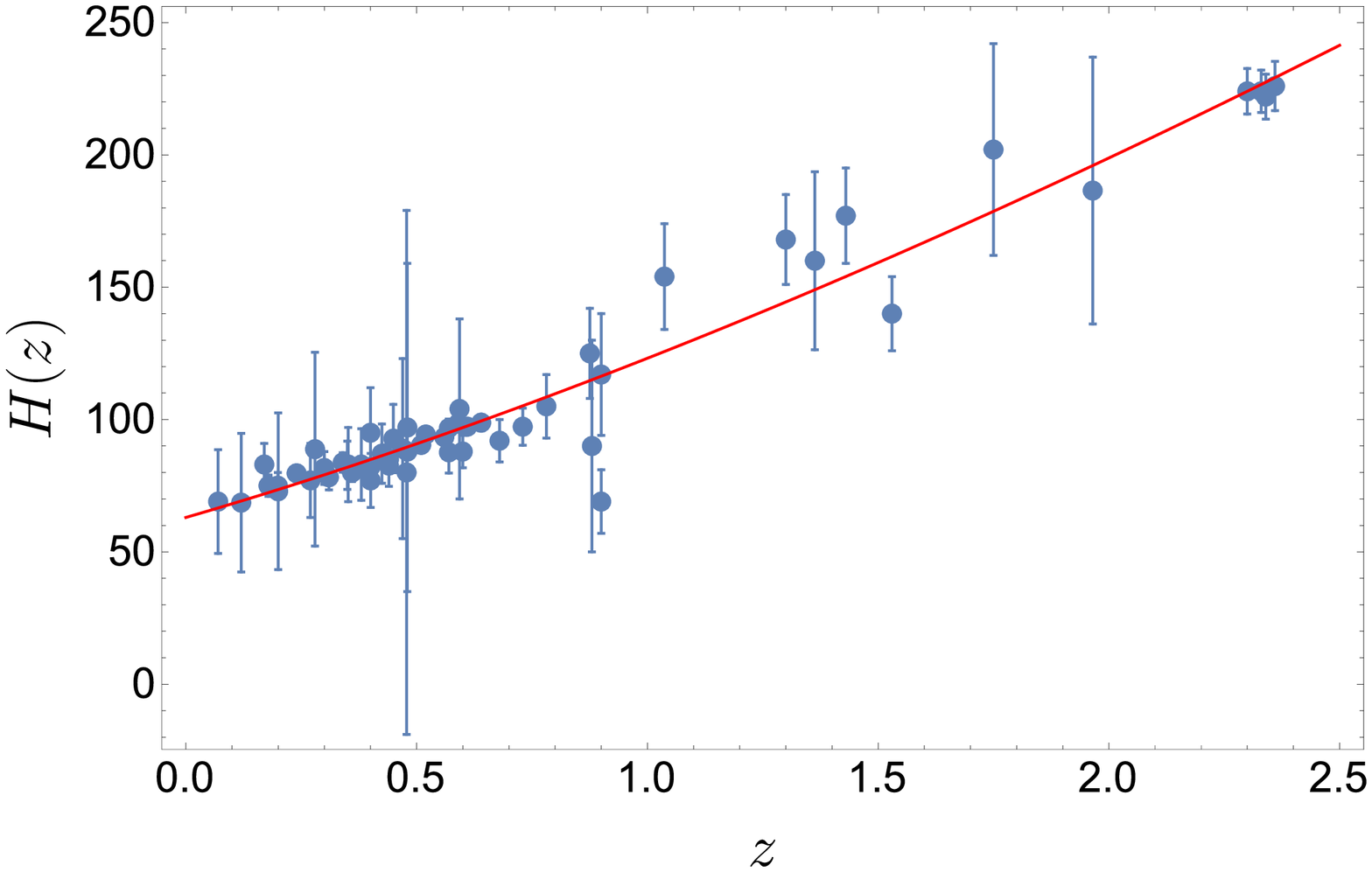}
\caption{Best fit curve of Hubble's parameter $H$ versus redshift $z$. Dots represent observed values, while the red curve is the theoretical fit.}
\label{PlotH}
\end{center}
\end{figure}

From the fit in Fig.~\eqref{PlotH}, we can also
infer the current value of Hubble's parameter predicted by the present model as $H_0=(63.3\pm4.3)\,\mathrm{km\,s^{-1}\,Mpc^{-1}}$. This is consistent with the recent observation from Planck Collaboration $H_0=\left(67.27\pm0.60\right)\mathrm{km\,s^{-1}\,Mpc^{-1}}$~\cite{Planck}, 
but still deviates from the value $H_0=\left(74.03\pm1.42\right)\mathrm{km\,s^{-1}\,Mpc^{-1}}$ based on the analysis of 
2019 SH0ES collaboration~\cite{SH0ES}. More discussion on this
point can be found in the concluding section.

\begin{table}[h!]
  \centering
    \begin{tabular}{|c|c|c|c|c|c|}
    \hline
  \,  $z$\, &\, $H(z)\,$\, &\, $\sigma_H$\, & \,$z\,$ & \,$H(z)$\,& \, $\sigma_H$\,\\
  \hline
  \hline
  \,0.070\, &\, 69.0\, &\, 19.6\, &\, 0.4783\, &\hspace{-0.4mm} 80\, &\, 99\,\\
  \hline
0.90 & 69 & 12 & 0.480 & 97 & 62 \\
\hline
0.120 & 68.6 & 26.2 & 0.593 & 104 & 13 \\
\hline
0.170 & 83 & 8 & 0.6797 & 92 & 8\\
\hline
0.1791 & 75 & 4 & 0.7812 & 105 & 12\\
\hline
0.1993 & 75 & 5 & 0.8754 & 125 & 17\\
\hline
0.200 & 72.9 & 29.6 & 0.880 & 90 & 40\\
\hline
0.270 & 77 & 14 & 0.900 & 117 & 23\\
\hline
0.280 & 88.8 & 36.6 & 1.037 & 154 & 20\\
\hline
0.3519 & 83 & 14 & 1.300 & 168 & 17 \\
\hline
0.3802 & 83.0 & 13.5 & 1.363 & 160.0 & 33.6\\
\hline
0.400 & 95 & 17 & 1.430 & 177 & 18\\
\hline
0.4004 & 77.0 & 10.2 & 1.530 & 140 & 14\\
	\hline
0.4247 & 87.1 & 11.2 & 1.750 & 202 & 40\\
\hline
0.4497 & 92.8 & 12.9 & 1.965 & 186.5 & 50.4\\
\hline
0.470 & 89 & 34 & & & \\
\hline
      \end{tabular}
      
      \vspace{4mm}
      \begin{tabular}{|c|c|c|c|c|c|}
      \hline
  \,  $z$\, &\, $H(z)\,$\, &\, $\sigma_H$\, & \,$z\,$ & \,$H(z)$\,& \, $\sigma_H$\,\\
  \hline
  \hline
  \,0.24\, &\, 79.69\, &\, 2.99\, &\, 0.52\, &\, 94.35\, &\, 2.64\,\\
  \hline
0.30 &\, 81.70\, & 6.22 & 0.56 & 93.34 & 2.30\\
\hline
0.31 & 78.18 & 4.74 & 0.57 & 87.6 & 7.8\\
\hline
0.34 & 83.80 & 3.66 & 0.57 & 96.8 & 3.4\\
\hline
0.35 & 82.7 & 9.1 & 0.59 & 98.48 & 3.18\\
\hline
0.36 & 79.94 & 3.38 & 0.60 & 87.9 & 6.1\\
\hline
0.38 & 81.5 & 1.9 & 0.61 & 97.3 & 2.1\\
\hline
0.40 & 82.04 & 2.03 & 0.64 & 98.82 & 2.98\\
\hline
0.43 & 86.45 & 3.97 & 0.73 & 97.3 & 7.0\\
\hline
0.44& 82.6& 7.8& 2.30& 224.0& 8.6\\
\hline
0.44& 84.81& 1.83& 2.33& 224& 8\\
\hline
0.48 & 87.90 & 2.03 & 2.34 & 222.0 & 8.5\\
\hline
0.51 & 90.4 & 1.9 & 2.36 & 226.0 & 9.3\\
\hline
      \end{tabular}
  \caption{57 experimental points of $H(z)$: the first 31 have been obtained from the method of Differential Age, while the last 26 from BAO and other approaches ($H$ is expressed in $\mathrm{km\,s^{-1}\,Mpc^{-1}}$ and $\sigma_H$ represents the uncertainty
for each data point).}
  \label{TabI}
\end{table}

\section{Thermal stability of Barrow Holographic Dark Energy}
\label{Thermal}
Let us now study our model from the 
thermodynamic point of view. Toward this end, 
we follow the analysis of~\cite{Abdollah,Bhandari}.
In particular, 
we are going to investigate whether
the thermal stability criteria are satisfied. 

We start by defining the internal
energy of the cosmic fluid (dark energy + dark matter) 
in the Universe as
\be
\label{toten}
E=V\left(\rho_D+\rho_M\right)\equiv V\rho_T\,,
\ee
where $V$ and $\rho_T$ 
are the volume 
and total energy density
of the cosmic system, respectively. 
We can then write the thermodynamic relation~\cite{Callen}
\be
\label{TS}
T S = E + p_D V\,,
\ee
where $T$ denotes the temperature of the cosmic system and 
we have implicitly taken into account that DM
is pressureless (i.e. $p_M=0$). 

\begin{figure}[t]
\begin{center}
\includegraphics[width=8.5cm]{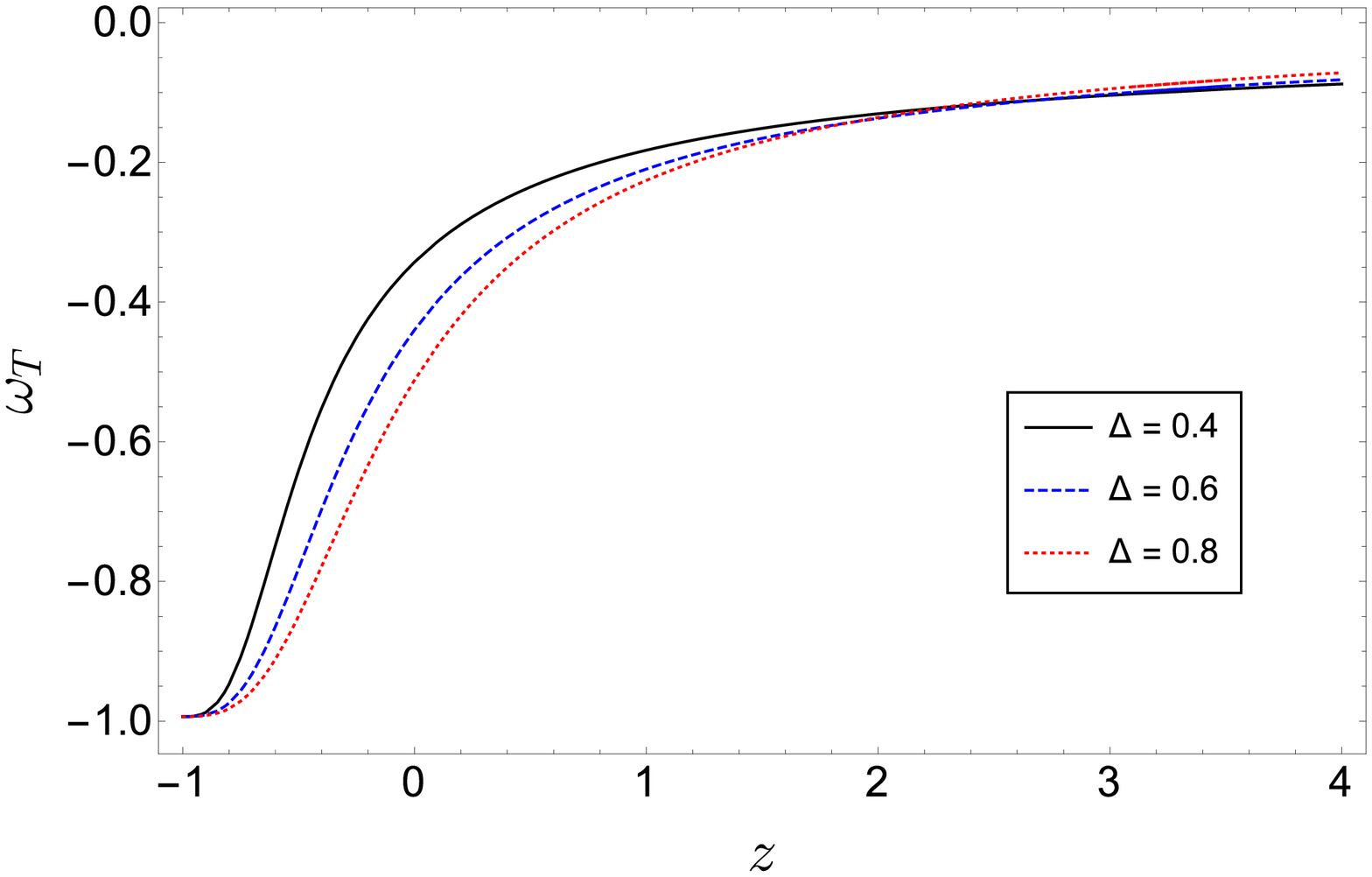}
%\caption{The evolution of $\Omega_D$ versus $z$ for different values
%of $b^2$. We have set $\Omega_k=0.01$, $\Delta=0.4$ and $\Omega_D^0=0.73$ as initial condition.}
%\label{Fig6}
\end{center}
\begin{center}
\includegraphics[width=8.5cm]{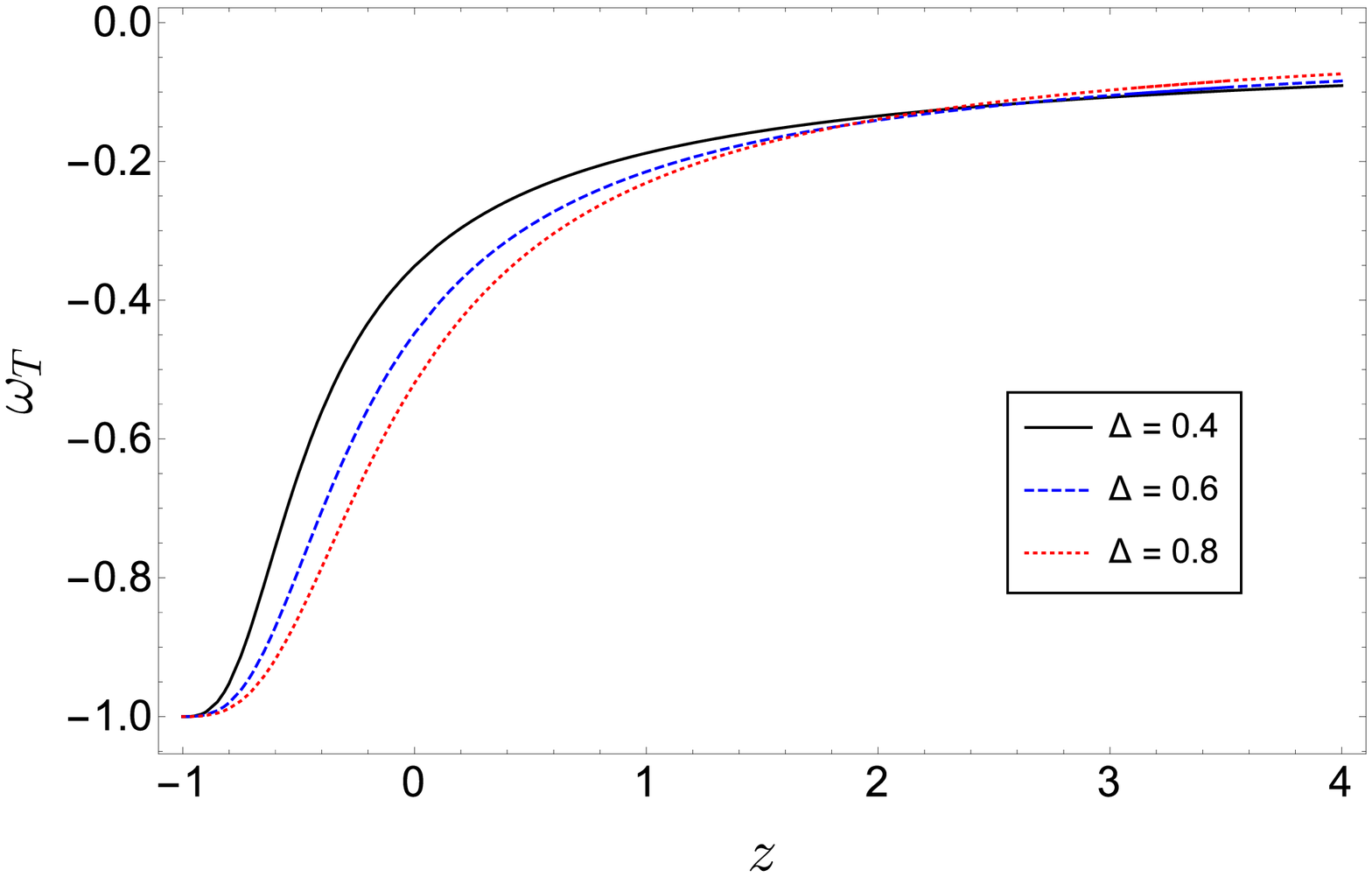}
\caption{The evolution of $\omega_T$ versus $z$ for noninteracting
BHDE model and different values
of $\Delta$. We have set $\Omega_D^0=0.73$ as initial condition and $\Omega_k=0.01$ in the upper panel, while $\Omega_k=0$ in the lower one.}
\label{Fig11}
\end{center}
\end{figure}

Inverting Eq.~\eqref{TS} respect
to $E$ and using Eq.~\eqref{toten}, we get
\be
\label{ESrel}
E= \frac{S\,T}{1+\omega_T}\,,
\ee
where $\omega_T\equiv p_D/\rho_T$ is the total
EoS parameter of the cosmic fluid. 
The behavior
of $\omega_T$ versus $z$ is plotted in Fig.~\ref{Fig11} and
Figs.~\ref{Fig13}-\ref{Fig14} for noninteracting and interacting BHDE models, respectively. It can be seen that  
the evolution trajectories of this parameter 
are not considerably affected 
by either the curvature $\Omega_k$ or the 
interaction term $b^2$.

\begin{figure}[t]
\begin{center}
\includegraphics[width=8.5cm]{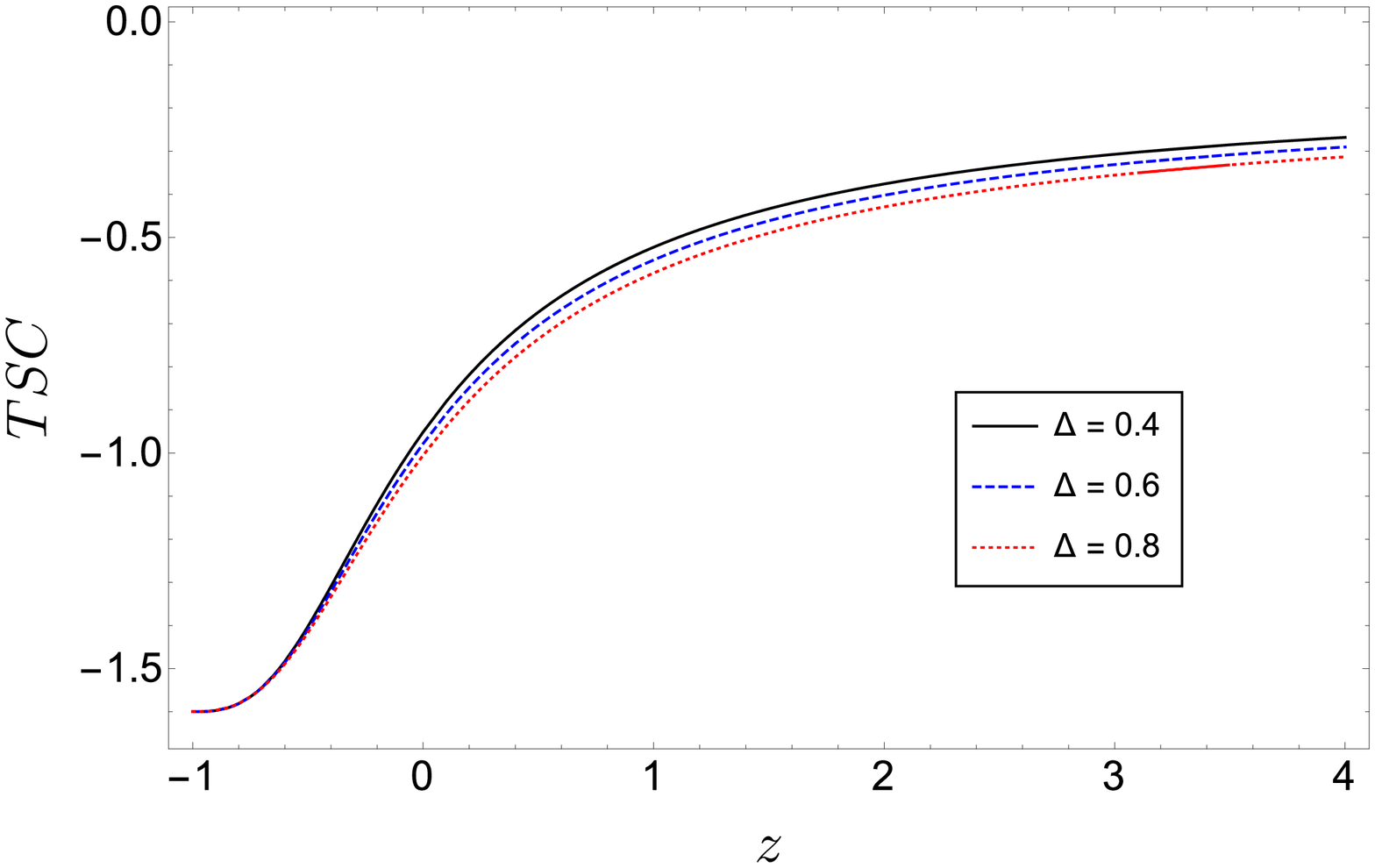}
%\caption{The evolution of $\Omega_D$ versus $z$ for different values
%of $b^2$. We have set $\Omega_k=0.01$, $\Delta=0.4$ and $\Omega_D^0=0.73$ as initial condition.}
%\label{Fig6}
\end{center}
\begin{center}
\includegraphics[width=8.5cm]{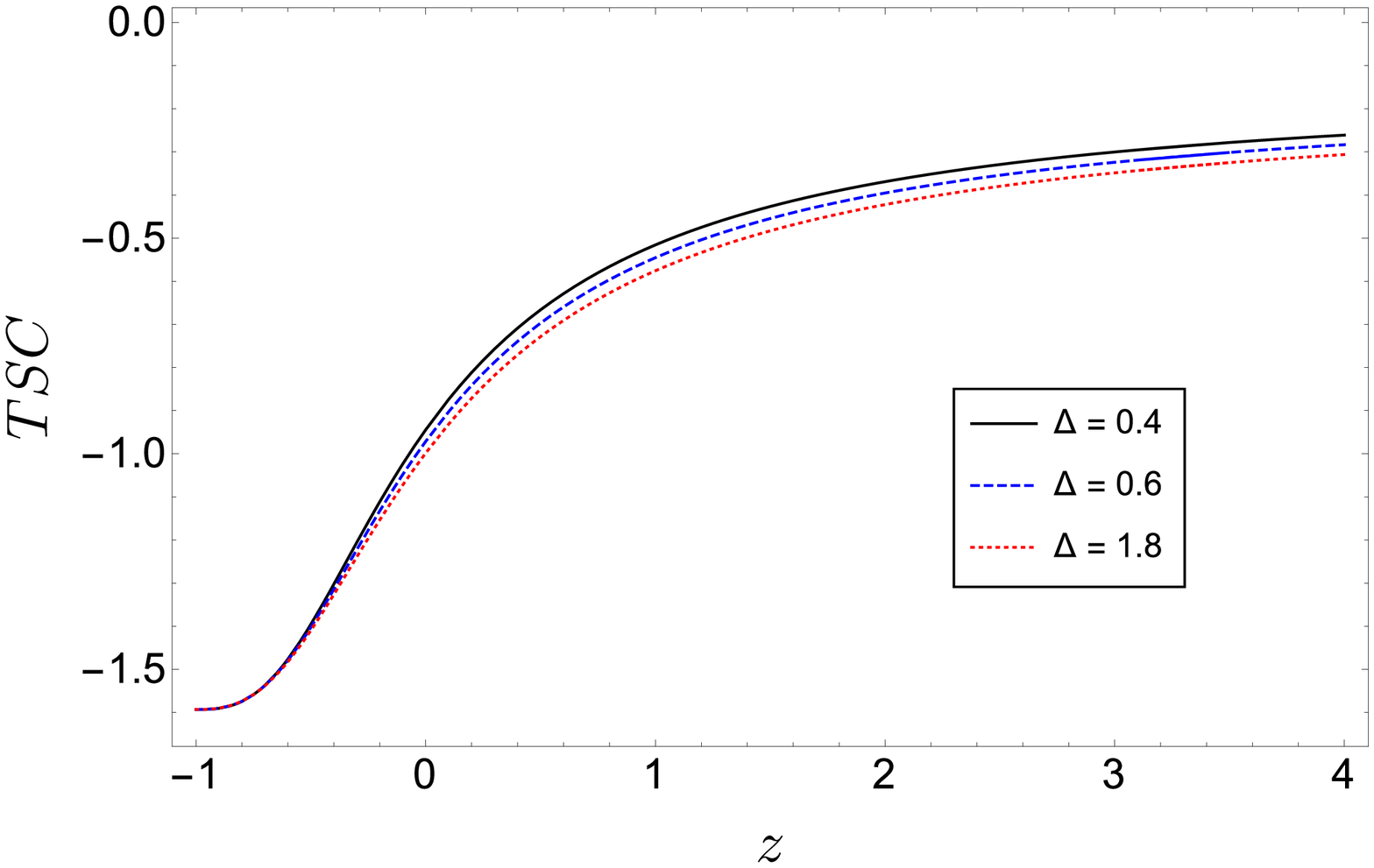}
\caption{Thermal stability condition (TSC) for noninteracting BHDE. We have set $\Omega_D^0=0.73$ as initial condition and $\Omega_k=0.01$ in the upper panel, while $\Omega_k=0$ in the lower one.}
\label{Fig12}
\end{center}
\end{figure}

\begin{figure}[t]
\begin{center}
\includegraphics[width=8.5cm]{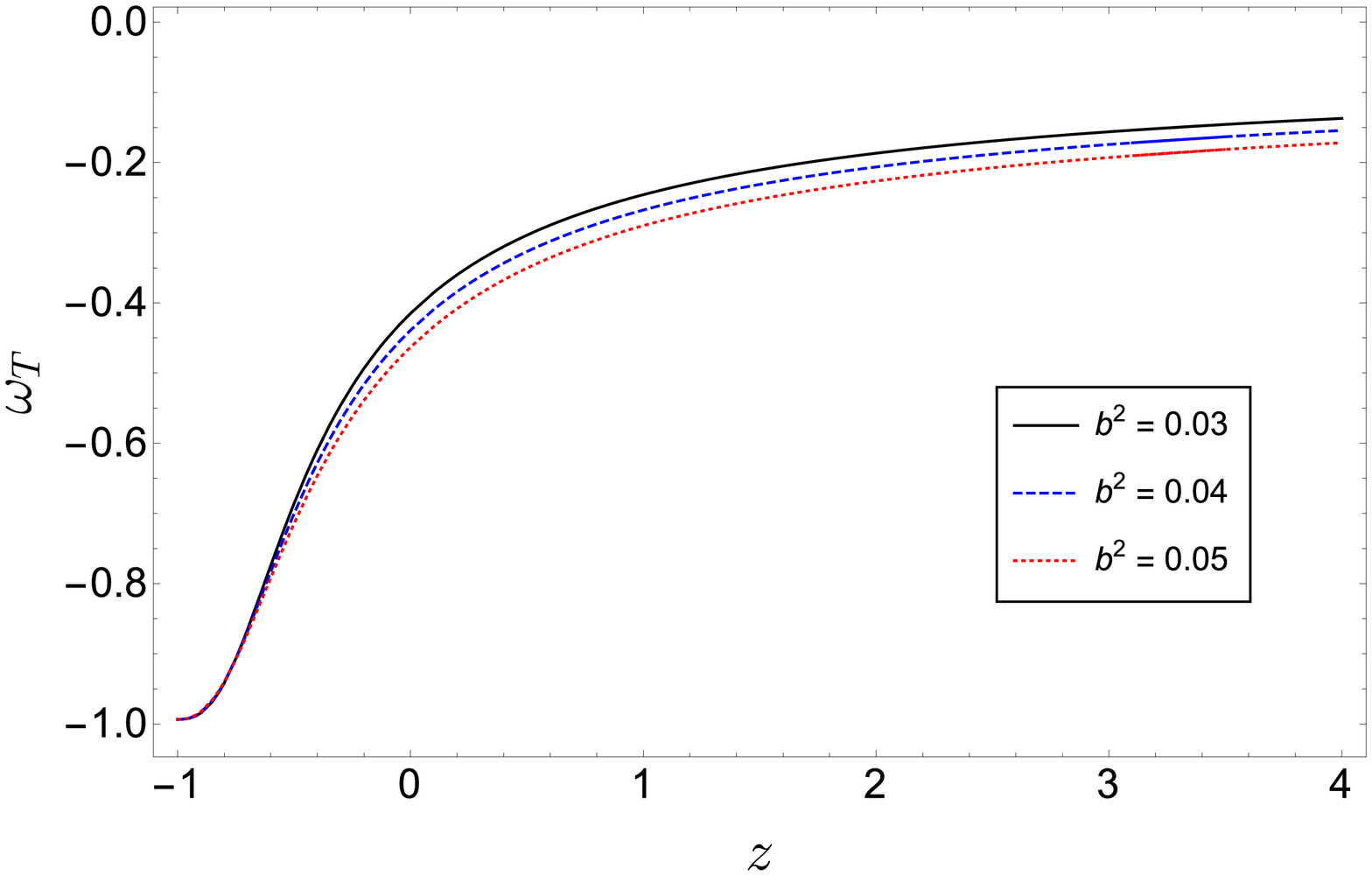}
%\caption{The evolution of $\Omega_D$ versus $z$ for different values
%of $b^2$. We have set $\Omega_k=0.01$, $\Delta=0.4$ and $\Omega_D^0=0.73$ as initial condition.}
%\label{Fig6}
\end{center}
\begin{center}
\includegraphics[width=8.5cm]{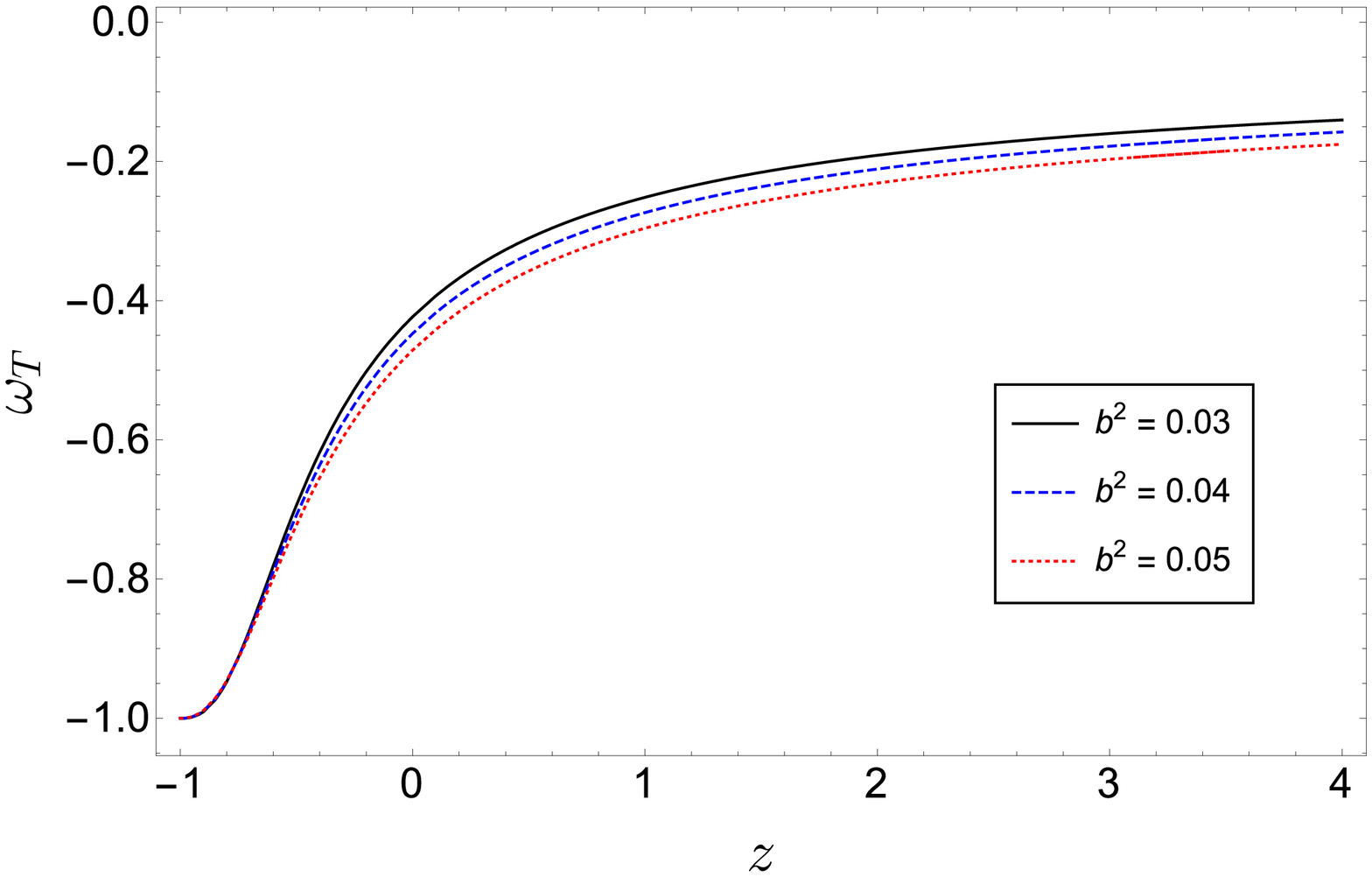}
\caption{The evolution of $\omega_T$ versus $z$ for interacting
BHDE model and different values of $b^2$. We have set $\Delta=0.4$ and $\Omega_D^0=0.73$ as initial condition. In the upper panel we have considered $\Omega_k=0.01$, while in the lower one $\Omega_k=0$.}
\label{Fig13}
\end{center}
\end{figure}

\begin{figure}[t]
\begin{center}
\includegraphics[width=8.5cm]{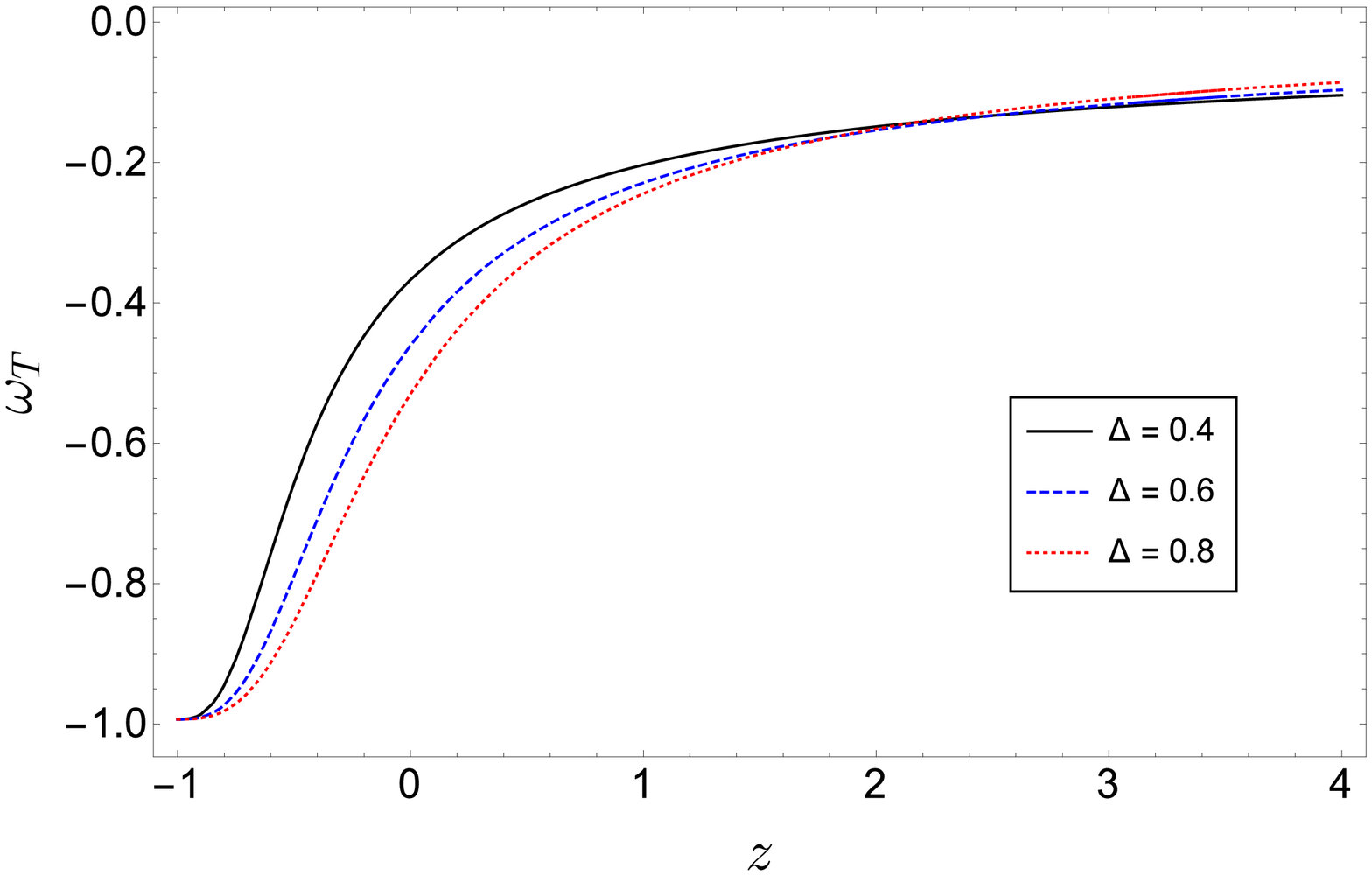}
%\caption{The evolution of $\Omega_D$ versus $z$ for different values
%of $b^2$. We have set $\Omega_k=0.01$, $\Delta=0.4$ and $\Omega_D^0=0.73$ as initial condition.}
%\label{Fig6}
\end{center}
\begin{center}
\includegraphics[width=8.5cm]{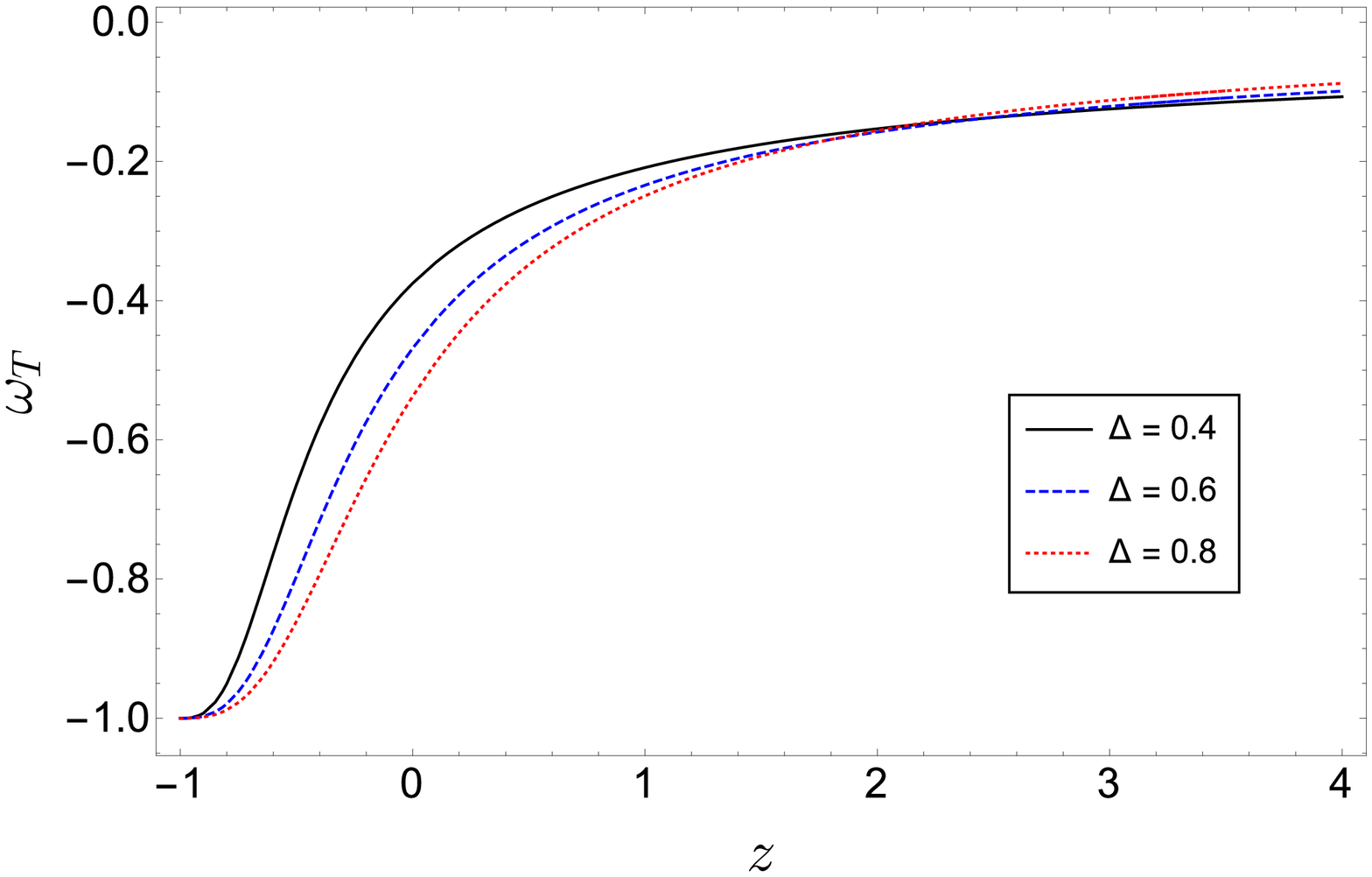}
\caption{The evolution of $\omega_T$ versus $z$ for interacting
BHDE model and different values of $\Delta$. We have set $b^2=0.01$ and $\Omega_D^0=0.73$ as initial condition. In the upper panel we have considered $\Omega_k=0.01$, while in the lower one $\Omega_k=0$.}
\label{Fig14}
\end{center}
\end{figure}

\begin{figure}[t]
\begin{center}
\includegraphics[width=8.5cm]{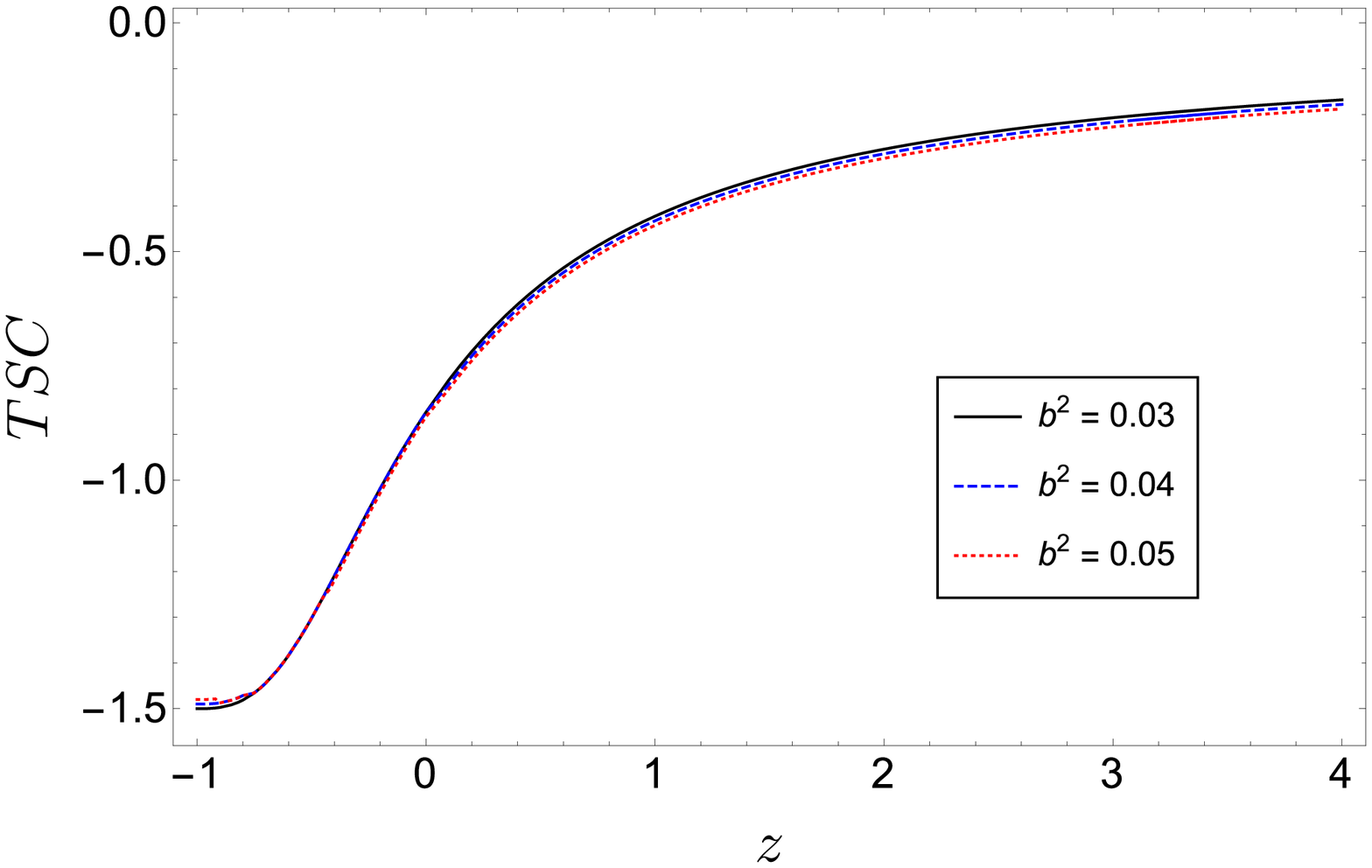}
%\caption{The evolution of $\Omega_D$ versus $z$ for different values
%of $b^2$. We have set $\Omega_k=0.01$, $\Delta=0.4$ and $\Omega_D^0=0.73$ as initial condition.}
%\label{Fig6}
\end{center}
\begin{center}
\includegraphics[width=8.5cm]{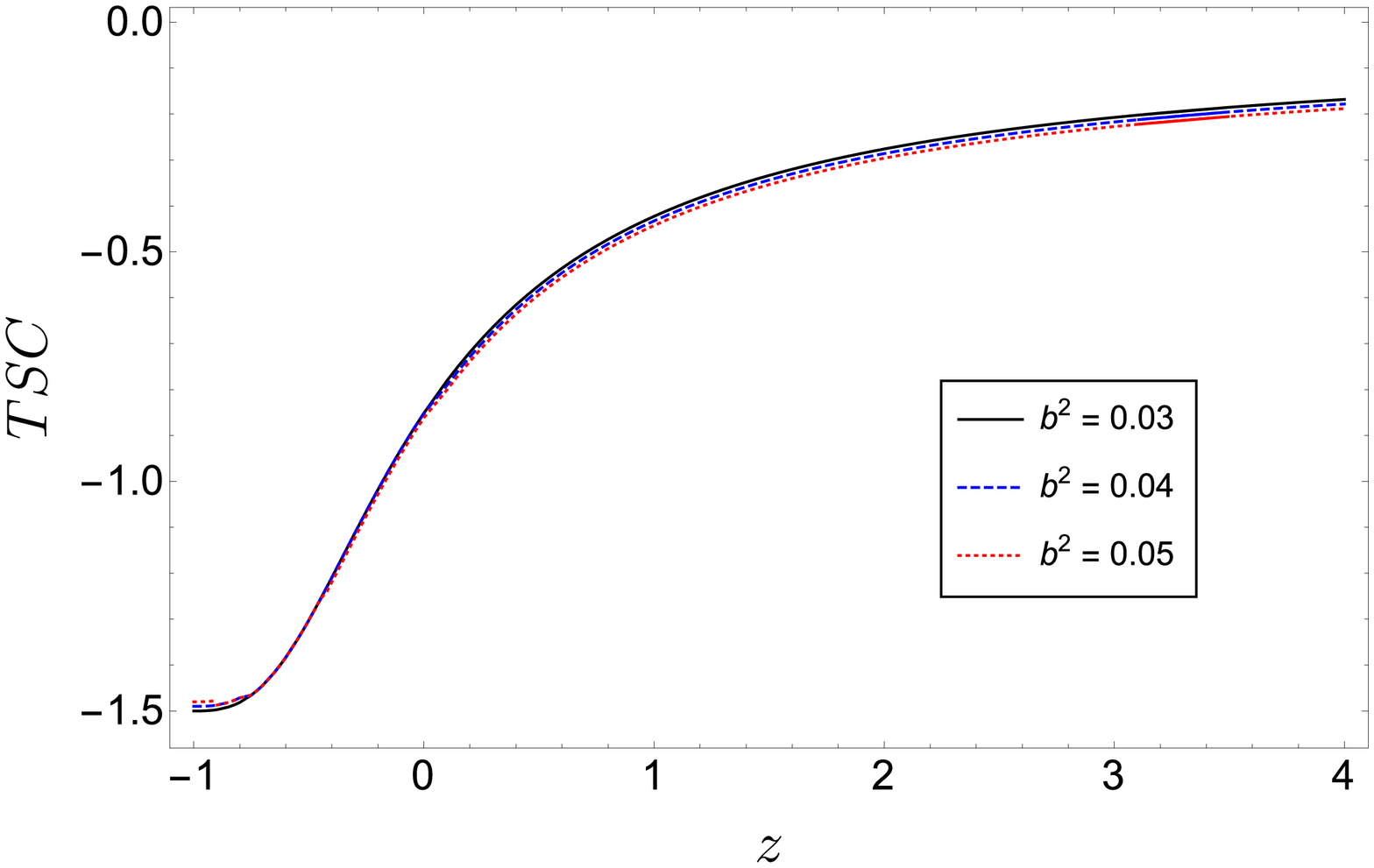}
\caption{Thermal stability condition (TSC) for interacting BHDE and different values of $b^2$. We have set $\Delta=0.4$ and $\Omega_D^0=0.73$ as initial condition. In the upper panel we have considered $\Omega_k=0.01$, while in the lower one $\Omega_k=0$.}
\label{Fig15}
\end{center}
\end{figure}

Let us now consider a reversible adiabatic expansion
($dS=0$). By using the first law of thermodyanmic
combined with Eq.~\eqref{ESrel}, we are led to~\cite{Bhandari}
\be
\label{E2}
E=E_0\left(\frac{1+\omega_0}{1+\omega_T}\right)\frac{T}{T_0}\,,
\ee
where $E_0$ and $T_0$ are the present values
of the cosmic fluid energy and temperature, respectively, 
while $\omega_0$ is the present value of the total EoS
parameter. 
Hence, comparison of Eqs.~\eqref{ESrel} and~\eqref{E2} gives 
\be
S= \frac{\left(1+\omega_0\right)E_0}{T_0}=const.
\ee

Additionally, using the heat capacities and the total
energy-momentum conservation law, we obtain~\cite{Barboza,Bhandari}
\begin{eqnarray}
\label{Cv}
\hspace{-4mm}C_v&=&\left(\frac{\partial E}{\partial T}\right)_V=\frac{C_p\, d\log V}{\left(1+\omega_T\right)d \log V-d \log |\omega_T|}\,,\\[2mm]
\hspace{-4mm}C_p&=&\left(\frac{\partial h}{\partial T}\right)_p=\frac{\left(1+\omega_T\right)E}{T}=\frac{\left(1+\omega_0\right)E_0}{T_0}=S\,,
\label{Cp}
\end{eqnarray}
where we have resorted to Eq.~\eqref{E2} in the third
step of the second relation. Here, $C_v$ and $C_p$ 
are the heat capacities at constant volume and pressure, 
respectively, and 
\be
h=E+p\,V\,,
\ee 
is the enthalpy of the cosmic fluid.

By combining Eqs.~\eqref{Cv} and~\eqref{Cp}, one can show that
\be
\label{Cpbis}
C_p=\left(1+\omega_T-\frac{d\log|\omega_T|}{d\log V}\right)C_v\,.
\ee
Let us also introduce the definitions of thermal
expansivity, isothermal compressibility and adiabatic compressibility.
They are given by
\begin{eqnarray}
\label{alpha1}
\alpha&=&\frac{1}{V}\left(\frac{\partial V}{\partial T}\right)_p\,,\\[2mm]
\label{kt}
\kappa_T&=&-\frac{1}{V}\left(\frac{\partial V}{\partial p}\right)_T\,,\\[2mm]
\label{ks}
\kappa_S&=&-\frac{1}{V}\left(\frac{\partial V}{\partial p}\right)_S\,,
\end{eqnarray}
respectively. 
By use of Eqs.~\eqref{Cv}-\eqref{alpha1},  
we then have~\cite{Abdollah}
\be
\label{alpha}
\alpha=\frac{C_v}{V\rho_T\,\omega_T}\left(\omega_T-\frac{d\log|\omega_T|}{d\log V}\right).
\ee
Moreover, the following relations between the heat capacities
$\kappa_T$ and $\kappa_S$ hold
\begin{eqnarray}
\label{cond1}
&&\kappa_T=\frac{\alpha V}{C_p}\,,\qquad \kappa_S=\frac{\alpha V C_v}{C_p^2}\,,\\[2mm]
&&\hspace{13.5mm} \frac{\kappa_T}{\kappa_S}=\frac{C_p}{C_v}\,.
\label{cond2}
\end{eqnarray}

We now analyze the thermal stability 
of the present system assuming that work
is done due to the volume variation of the thermal system only. 
For stable equilibrium, it has been shown that the second order
variation of the internal energy
\be
\label{secord}
\delta^2E=\delta T\,\delta S-\delta p\,\delta V\,,
\ee
must satisfy the condition $\delta^2E\ge0$~\cite{Callen}. 

Notice that, by using 
Eqs.~\eqref{Cv},~\eqref{Cp},~\eqref{kt} and~\eqref{ks}, 
we can equivalently cast Eq.~\eqref{secord} as
\be
\delta^2E=V\,\kappa_S\,\delta p^2+\frac{T}{C_p}\delta S^2\,,
\ee
or
\be
\delta^2E=\frac{1}{V\hspace{0.3mm}\kappa_T}\,\delta V^2+\frac{C_v}{T}\,\delta T^2\,.
\ee
Thus, for the stability of the system, the 
heat capacities and compressibilities should obey
\be
\label{alsat}
C_p, C_v, \kappa_S, \kappa_T \ge0\,, 
\ee
which, combined with Eqs.~\eqref{cond1} and~\eqref{cond2},
leads to~\cite{Abdollah}
\be
\label{inequa}
C_p\ge C_v\,,\qquad 
\kappa_T\ge \kappa_S\,.
\ee

From Eq.~\eqref{Cp}, it is straightforward to see 
that $C_p$ is constant and positive-definite. Moreover, Eqs.~\eqref{Cpbis}
and~\eqref{alpha} imply that the constraints~\eqref{inequa}
and, thus,~\eqref{alsat} are met only if the total EoS parameter $\omega_T$ obeys the Thermal Stability Condition (TSC)
\be
\label{TSC}
\omega_T-\frac{d\log|\omega_T|}{d\log V}\ge 0\,.
\ee
In Fig.~\ref{Fig12} and Fig.~\ref{Fig15} 
the TSC has been plotted
for the noninteracting and interacting cases, respectively. One can 
see that such criterion is not satisfied, 
indicating that the cosmic fluid
is not thermodynamically stable in BHDE model.

\section{Conclusions and Outlook}
\label{Conc}
We have analyzed the cosmic evolution of a nonflat
FRW Universe filled by pressureless
dark matter and BHDE. By considering
the apparent horizon as IR cutoff, we have examined
the evolutionary trajectories of various model parameters, 
such as the BHDE density parameter, the EoS parameter, 
the deceleration parameter, the jerk parameter and the squared sound speed. 
In the absence of interaction, we have found that 
$\omega_D$ never crosses the phantom line. 
Furthermore, 
the model successfully predicts the 
sequence of an early
matter dominated era with a decelerated expansion, followed by a late time DE dominated
epoch with an accelerated phase. 
Nevertheless, 
the values of the current EoS parameter, transition redshift and deceleration parameter are inconsistent 
with recent results in the literature (see Table~\ref{TabII}). 

On the other hand,  assuming 
a mutual interaction between the dark sectors of the cosmos, 
we have obtained that BHDE enters the phantom
regime in the future, with such a behavior being
driven by the coupling $b^2$. The model still predicts
the usual thermal history of the Universe
and is consistent with observations for what concerns
the values of the EoS parameter, transition redshift and deceleration parameter in the current epoch.

\begin{widetext}
 \begin{center}
\begin{table}[h]
    \begin{tabular}{|c|c|c| c|}
    \hline
 &\,\,Non-interacting model \,\, &\,\, Interacting model  \,\,&\,\, Observational value\,\, \\
 \hline
 \hline
$\omega_0$ \,\, &\,\, $[-0.72,-0.47]$ &\,\,  $[-0.96,-0.74]$ \,\,&\,\, $[-1.38,-0.89]$\, (Planck+WP+BAO) \,\, \\
           \hline
           $z_t$ \,\, &\,\, $[0.02,0.46]$ &\,\,  $[0.30,0.64]$ \,\,&\,\, $0.61$\, (SNIa+CMB+LSS)\,\,\\
           \hline
           $q_0$ \,\, &\,\, $[-0.26,-0.01]$ &\,\,  $[-0.33,-0.03]$ \,\,&\,\, $[-0.86,-0.42]$\, (Union2 SNIa)\,\,\\
           \hline
            $j_0$ \,\, &\,\, $[0.36,0.62]$  &\,\,  $[0.35,0.64]$  \,\,&\,\, $j=1$ ($\Lambda$CDM) \,\,\\
            \hline
            $H_0$ \,\, &\,\, $-$  &\,\,  $(63.3\pm4.3)\,\mathrm{km\,s^{-1}\,Mpc^{-1}}$  \,\,&\,\, $\left(67.27\pm0.60\right)\mathrm{km\,s^{-1}\,Mpc^{-1}}$ (Planck)\,\,\\
            \hline
    \end{tabular}
  \caption{Theoretical and observational values of EoS parameter, transition redshift, deceleration parameter, jerk parameter and Hubble parameter in the current epoch (for the jerk parameter we have considered the $\Lambda$CDM prediction as reference value).}
  \label{TabII}
\end{table}
\end{center}
\end{widetext}

Furthermore, for the interacting model
we have studied the evolution
of $H(z)$  and compared it
with the data points obtained from the 
latest compilation of 57 Hubble's parameter measurements from Differential Age, BAO and other approaches. By using the $R^2$-test, we have found the best fit of $H(z)$ 
for the model parameters $b=0.12$ and $\Delta=0.08$
(with $95\%$ confidence bounds). Also, the current value of Hubble
parameter turns out to be
$H_0=(63.3\pm4.3)\,\mathrm{km\,s^{-1}\,Mpc^{-1}}$, which is consistent with the recent constraint from Planck Collaboration~\cite{Planck}, but is in tension
with the observation from 2019 SH0ES Collaboration~\cite{SH0ES}. 
Along this line, it could be interesting to explore whether 
BHDE (or, more general, HDE based on deformed entropies) can
play some role in solving the $H_0$
tension~\cite{H0Tension}.
However, a detailed investigation of this issue would require
the additional incorporation of CMB data and the development
of a joint analysis. This aspect goes beyond the scope of our
analysis and will be addressed
in future works.

We have finally studied the thermal stability of our model
by considering the heat capacities
and compressibilities. We have shown that both 
the noninteracting and interacting frameworks
suffer from the satisfaction of the Thermal Stability 
Condition~\eqref{TSC}. This result is in line with
the achievement of~\cite{Abdollah} for the case
of Tsallis Holographic Dark Energy and, more
general, with the outcome of~\cite{Barboza}, where
it has been found that DE fluids with a time-dependent 
EoS parameter are in
conflict with the physical constraints imposed by thermodynamics.

Further aspects are still to be considered: 
first, one could investigate how the
present study interfaces with the results of~\cite{Mamon:2018flx}, where it has been shown that some DE
models, such as Chevallier-Polarski-Linder model,  
Generalized Chaplygin Gas and
Modified Chaplygin Gas are thermodynamically stable
for certain values of the model parameters.
A potential path to explore is to 
see whether thermal stability for BHDE
is achieved by considering different IR cutoffs and/or
generalized interactions between the dark sectors
of the cosmos. Of course, it is possible that even after
such modifications the TSC is violated, indicating
that BHDE may not be the answer to the
question of the unknown DE.

Additionally, since our framework provides
a preliminary attempt to study quantum gravity
effects on the cosmic history of the Universe,
it is worth analyzing to what extent it is consistent with 
predictions of more fundamental candidate theories of quantum gravity.
Finally, we aim to examine in more detail
the connection between BHDE and other 
models proposed to explain the origin of dark energy~\cite{Yoo}, and possibly fix more stringent constraints on the deformation parameter $\Delta$. 
These lines of research are under
active investigation and are left for future projects.

\acknowledgments 

The author acknowledges Manos Saridakis for
useful discussions and the anonymous Referee for insightful
comments that helped to improve the quality of the manuscript. 
He is also grateful to the Spanish ``Ministerio de Universidades'' 
for the awarded Maria Zambrano fellowship and funding received
from the European Union - NextGenerationEU. He finally acknowledges 
participation in the COST Association Action CA18108  ``Quantum Gravity Phenomenology in the Multimessenger Approach''.

\end{document}